\documentclass[cup6a]{cupconf}
\RequirePackage{graphicx}%
\RequirePackage{epsf}%
\input{psfig.sty}

\title[The influence of WR stars in the ISM of galaxies]{Are Wolf-Rayet stars able
to pollute the interstellar medium of galaxies? Results from integral
field spectrocopy }

  \author[P\'erez-Montero et al.]{Enrique P\'erez-Montero$^1$, Carolina Kehrig$^1$, Jarle Brinchmann$^2$,
Jos\'e M. V\'\i lchez$^1$, Daniel Kunth$^3$, \& Florence Durret$^3$}

  \affiliation{
$^1$ Instituto de Astrof\'\i sica de Andaluc\' \i a - CSIC. Apdo. 3004, 18008, Granada, Spain \\
$^2$ Leiden Observatory. Leiden University, PO Box 9513, 2300 RA Leiden, The Netherlands\\
$^3$ Institut d'Astrophysique de Paris, UMR 7095 CNRS, Universit\'e  Pierre \& Marie Curie, 98 bis boulevard Arago, 75014 Paris, France\\}

\newcommand{\hii}{H\,{\sc ii}\rm}

\newcommand{\nii}{[N\,{\sc ii}]}
\newcommand{\oiii}{[O\,{\sc iii}]}
\newcommand{\oii}{[O\,{\sc ii}]}

\newcommand{\hbeta}{H$\beta$}

\newcommand{\apj}{ApJ}
\newcommand{\apjl}{ApJL}
\newcommand{\apjs}{ApJS}
\newcommand{\aap}{A\&A}
\newcommand{\mnras}{MNRAS}
\newcommand{\aj}{AJ}

\begin{document}
\maketitle

\begin{abstract}
We investigate the spatial distribution of chemical abundances in a sample of low metallicity 
Wolf-Rayet (WR) galaxies selected from the SDSS.
We used the integral field spectroscopy technique in the optical spectral 
range (3700 \AA-6850 \AA) with PMAS attached to the CAHA 3.5 m telescope.
Our statistical analysis of the spatial distributions of O/H and N/O, as derived using
the direct method or strong-line parameters consistent with it, indicates that
metallicity is homogeneous in five out of the six analysed objects in
scales of the order of several kpc. Only in the object WR404, a gradient of
metallicity is found in the direction of the low surface brightness tail. In contrast,
we found an overabundance of N/O in spatial scales of the order of hundreds of pc
associated with or close to the positions of the WR stars in 4 out of the
6 galaxies. We exclude possible
hydrodynamical causes, such as the metal-poor gas inflow, for this local pollution by means of the analysis of
the mass-metallicity relation (MZR) and mass-nitrogen-to-oxygen relation (MNOR) 
for the WR galaxies catalogued in the SDSS.

\end{abstract}

%

\section{Introduction}

Wolf-Rayet (WR) galaxies host very bright episodes of
star formation characterized by the emission of broad
WR bumps in their optical spectrum (Osterbrock \& Cohen, 1982). 
The two main bumps in the optical range are the blue bump, centered at a wavelength
of 4650 \AA, produced by the emission from
N{\sc v}, N{\sc iii}, C{\sc iii}/C{\sc iv} and He{\sc ii} and 
associated with WC and
WN stars, and the red bump which is fainter and is centered
at $\sim$ 5800 \AA, produced mainly by C{\sc iii} and C{\sc iv},
and associated with WC stars.
The lines making up
these bumps originate in the dense stellar winds from
WR stars ejecting metals into the interstellar medium
(ISM).

The number of known WR galaxies has tremendously increased 
from the discovery of the first one (He2-10: Allen et al. 1976), with
different published catalogs (Conti, 1991; Schaerer et al. 1999; 
Guseva et al. 2000; Zhang et
al. 2007) until the list of WR galaxies in the Sloan Digital Sky Survey (SDSS)
by Brinchmann et al. (2008) with around 570 objects with the identification
of the WR bumps in their integrated spectra.

There is an increasing evidence that the most challenging problems for this
kind of objects appears in the low metallicity galaxies. Although it is well
documented that the number of WR stars and the intensity of the WR bumps are
higher for higher metallicities (Crowther, 2007), the values found in some
low metallicity {\hii} galaxies, such as IZw18 (Legrand et al. 1997) are claimed to be much
higher than those predicted by synthesis population models ({\em e.g.} 
Gonz\'alez-Delgado et al. 2005).

Among the other important open issues regarding WR galaxies  is the 
chemical enrichment of the ISM surrounding
the stellar clusters where the WR stars are located. It is well known that there 
is an overabundance of the
N/O ratio found in some WR nebulae (e.g. V\'\i lchez \& Esteban 1991, Esteban \&
V\'\i lchez 1992, Fern\'andez-Mart\'\i n et al. 2012) and also in the ISM of
some WR galaxies, where the WR features are diluted ({\em e.g.}  
HS0837+4717, Pustilnik et al. 2004, NGC5253, L\'opez-S\'anchez et al. 2007, 
other {\hii} galaxies in H\"agele et al. 2006, 2008, {\em green pea} galaxies,
Amor\'\i n et al. 2012). Brinchmann et al. (2008) also showed that the median N/O ratio in WR
galaxies with EW(H$\beta$) $<$ 100 \AA, has
an excess of $\sim$ 25$\%$ in relation to the other star-forming
galaxies in the SDSS.  Chemical
evolution models do not predict high N/O values in these
low-metallicity galaxies (Moll\'a et al. 2006), as for 12+log(O/H) $<$ 8.2 most
of the N in the ISM has a primary origin and therefore its chemical
abundance does not depend on the metallicity of the gas, and the
expected N/O ratio for closed-box models has a constant value around 
log(N/O) $\approx$ -1.5.
However, many of these integrated observations do not allow us
to properly relate the excess in some chemical species with their WR content.

To investigate the issue, among others, of the possible connection between the presence of WR stars
and the chemical pollution of the surrounding ISM we have carried out a program to study 
metal-poor WR galaxies by means of integral field spectroscopy (IFS; Kehrig et al. 2008, 
P\'erez-Montero et  al. 2011, Kehrig et al. 2013). 
Integrated observations, such as long-slit or fibers, may fail to correlate the spatial location 
and distribution of WR features with respect to the physical conditions and the chemical
abundances of the ISM as derived from the optical emission-lines.
Thus, a two-dimensional analysis of the ionized
material in galaxies helps us to better understand the interplay
between the massive stellar population and the ISM.  For instance,
whether WR stars are a significant contributor to abundance
fluctuations on timescales of {\it t} $\sim$ 10$^{7}$ yr and to the
formation of high-ionization lines (e.g. He{\sc ii} 4686 \AA) are still
unsolved issues ({\em e.g.} Kehrig et al. 2011, Shirazi \& Brichmann, 2012)
that can be probed more precisely when applying IFS to nearby galaxies
(see Kehrig et al. 2013).

Thus far, the results coming from WR galaxies studied with IFS point to different
scenarios, depending on the relative position between the local or extended N and/or He enrichment
and the location(s) of the WR stars. L\'opez-S\'anchez et al. (2011)
claimed to find a local N overabundance associated with WR emission in IC-10.
A similar result is found by James et al. (2013b) for the blue compact
dwarf galaxy Haro 11.
Monreal-Ibero et al. (2012) also find in NGC5253 local peaks of high N/O
(see also Westmoquette et al. 2013)
, but only 
some of them are associated with WR emission, so this could be indicative of
a different timescale between the formation of the WRs and the mixing of the
ejected material with the surrounding ISM. A similar scenario is found
by Kehrig et al. (2013) in Mrk178, where only one out of three detected
WR clusters can be associated with an overabundance of N and He. Finally, in
P\'erez-Montero et al. (2011) the IFS study of N overabundant objects HS0837+4717 and Mrk930, 
also identified as WR galaxies, lead to high values of N/O in scales of more than 1 kpc, much 
beyond the power of the observed WR stars to pollute the ISM in these
scales and thus pointing to other hydrodynamical processes affecting
the chemistry of the gas in these galaxies, such as the infall of metal-poor
gas (K\"oppen \& Hensler, 2005). James et al. (2013a) also propose a similar scenario for their results
of IFS observations of the merging galaxy UM448.

In this work we extend the sample of low-metallicity WR galaxies studied
by means of IFS by six objects selected from the
WR galaxy catalog by Brinchmann et al. (2008).
The paper is organized as follows. In \S 2 we describe the observed
WR galaxies and we report the IFS
observations and data reduction. In \S 3 we present our results, 
including emission-line maps and derivation of oxygen and nitrogen chemical
abundances and their distributions in the observed fields of view. 
We also describe the measurement of the WR bumps in the observed galaxies.
In \S 4 we discuss our results about the chemical pollution of ISM versus WR stars 
the surrounding ISM. Finally we summarize our results and present
our conclusions in \S 5.

\section{Data acquisition and reduction}

\subsection{Object selection and observations}\label{spectra}


We obtained IFS data of a sample of six objects 
selected 
from the SDSS WR galaxy catalog by Brinchmann et al. (2008) following three criteria: 
i) galaxies should be associated with a WR index
of 2 (convincing WR features in the SDSS spectrum) or 3 (very clear features) 
in the catalog, ii) the main ionization source should be dominated by star formation 
as derived using diagnostic diagrams based on strong emission-lines (Baldwin et al. 1981), and 
iii) galaxies should have oxygen abundances lower than half the solar value [12+log(O/H) $\approx$ 8.4]
as derived from the direct method, These criteria were completed with two other observational 
conditions: i) to have sizes smaller than the field of view (FoV) of PMAS in lens array
mode 16'' $\times$ 16'' in order to cover the
entire galaxy in one single pointing, and ii) all objects were visible from the
CAHA observatory in the assigned dates at an airmass lower than 1.2.
The six
target WR galaxies are listed in Table~\ref{sample}. Column (1)
quotes the names of the objects from the catalog of Brinchmann et al. (2008). 
Column (2) shows the redshift of each galaxy. Columns (3) and (4) give the object
coordinates. Column (5) gives the WR index as done by Brinchmann et al.
(2008), column (6) gives other names, and finally column (7) shows the observing
date.

The data were acquired with the Integral Field Unit (IFU) Postdam Multi-Aperture
Spectrophotometer (PMAS), developed at the Astrophysikalisches
Institut Potsdam (Roth et al. 2005). PMAS is attached to the 3.5 m
Telescope at the CAHA Observatory (Spain).  The PMAS
spectrograph is equipped with 256 fibers coupled to a $16\times16$
lens array. Each fiber has a spatial sampling of
$1\hbox{$^{\prime\prime}$}\times1\hbox{$^{\prime\prime}$}$ on the sky
resulting in a FoV of
$16\hbox{$^{\prime\prime}$}\times16\hbox{$^{\prime\prime}$}$ collecting
square areas known as spaxels. 

We were awarded a service mode run on the nights 2009 June 22 - 25
(program F09-3.5-27).  In addition, we continued our program with
additional time on 2009
October 11th as part of the commissing run for the new PMAS CCD. 

\begin{table*}
\centering
\caption[The sample of Wolf-Rayet galaxies]{The sample of observed Wolf-Rayet galaxies
with different properties, including names, redshifts, positions, WR index, and
date of observations with PMAS.}
\label{sample}
\begin{tabular} {lcccccc}
\hline
Name     & redshift   &   R.A (2000)      &   $\delta$(2000)       &  WR class & other designation & observing date \\  \hline
WR 038   & 0.0158    &  17h29m06.55s     & +56d53m19.23s           &   2 & SHOC 575  & 22-23 Jun 2009\\
WR 039   & 0.0472    &  17h35m01.24s     & +57d03m08.55s          &  2 &SHOC 579  & 25 Jun 2009 \\
WR 057   & 0.0179    &  00h32m18.59s     & +15d00m14.16s          &  3 &SHOC 022 & 11 Oct 2009 \\
WR 266   & 0.0213    &  15h38m22.00s     & +45d48m07.02s          &  2 & & 24-25 Jun 2009 \\
WR 404   & 0.0220    &  21h34m37.80s     & +11d25m10.19s          &  2  & CGCG 427-004 & 24-25 Jun 2009 \\
WR 505   & 0.0164    &  16h27m51.17s     & +13d35m13.73s          &  2 &  & 22-23 Jun 2009 \\ 
\hline
\end{tabular}
\end{table*}


During observations taken in 2009 June with the old PMAS 2K $\times$ 2K CCD, 
we used the V600 grating in two separate spectral ranges: the blue side,
covering a spectral range $\sim$ 3700-5200\ {\AA} (centered at 4500 \AA)
and the red one (centered at 6325 \AA), providing a spectral range
from $\sim$ 5350 to 6850~\AA. For the galaxy WR057, taken
with the new 4K $\times$ 4K PMAS CCD but with the same resolution, we were able to cover the whole optical
spectral range ($\sim$ 3700 - 6850 \AA) in one shot using the same
V600 grating. The data were binned by a factor of 2 in the spectral 
direction, yielding a spectral resolution of $\sim$ 1.6 \AA/pixel. 
The data were acquired under
non-photometric conditions and with a seeing varying between
1$^{\prime\prime}$ and 1$^{\prime\prime}$.5 . To avoid major
differential atmospheric refraction (DAR) effects, all expositions
were taken at air mass $<$ 1.2.  We used one single pointing for
each galaxy, covering in all cases the most intense burst of star
formation and its surroundings. Observations of the
spectrophotometric standard stars BD+253941 and PG1633 in the first run,
and BD+284211 for the calibration of WR057 in the second one, were obtained
during the observing nights for flux calibration. Bias frames, arc
exposures (HgNe), and spectra of a continuum lamp were taken following
the science exposures as part of the PMAS baseline calibrations.



\subsection{Data reduction}



For all objects observed with the PMAS 2K $\times$ 2K CCD, the first steps of
the data reduction were done through the R3D package (S\'anchez 2006).
We used the P3d tool (Sandin 2010) to perform the basic data reduction of WR057,
taken with the new PMAS 2K $\times$ 4K CCD. This CCD is read out in four
quadrants which have slightly different gains (Kelz et al. 2006). At
the time we observed WR057, P3d was the only software capable to
handle the characterists of the new CCD.

After trimmimg, combining the four quadrants for WR057 and bias
subtraction, the expected locations of the spectra were traced on a
continuum-lamp exposure obtained before each target exposure. We
extracted the target spectra by adding the signal from the 5 pixels
around the central traced pixel (that is the total object spectrum
width). With exposures of HgNe arc lamps obtained immediately after
the science exposures, the spectra were wavelength calibrated.

Fibers have different transmissions that may depend on the
wavelength. The continuum-lamp exposures were used to determine the
differences in the transmission fiber-to-fiber and to obtain a
normalized fiber-flat image, including the wavelength dependence. This
step was carried out by running the \textsc{fiber-flat.pl} script from
the R3D package.  In
order to homogenize the response of all the fibers, we divided our
wavelength calibrated science images by the normalized fiber-flat
(S\'anchez 2006). Then, to remove cosmic rays, different
exposures taken at the same pointing were combined using the
\textsc{imcombine} routine in IRAF\footnote{IRAF is distributed by
the National Optical Astronomical Observatories, which are operated
by the Association of Universities for Research in Astronomy, Inc.,
under cooperative agreement with the National Science Foundation.}.
Flux calibration was performed using the IRAF tasks \textsc{standard},
\textsc{sensfunc} and \textsc{calibrate}. We co-added the spectra of
the central fibers of the standard star to create a one-dimensional
spectrum that was used to obtain the sensitivity function.  

The reduced spectra were contained in a data cube for each object.

\section{Results}

\subsection{Line measurements}

The emission-line fluxes on the extracted one-dimensional spectra were measured for each spaxel using 
a Gaussian fitting  over the local position of the continuum. This procedure was done
using an automatic routine based on the IRAF task \textsc{splot}. In the
case of those lines with a lower signal-to-noise (S/N) ({\em e.g.} \oiii~4363 {\AA} and
\nii~6584 \AA), the results
from this routine were revised by eye inspection and, if necessary, repeated using a
manual measurement.

The adjacent continuum to each line can be affected by the underlying stellar
population which can depress the intensity of the Balmer emission lines
with stellar absorption wings ({\em e.g.} Diaz 1988). This stellar absorption
was studied for the brightest spaxels by fitting a combination of synthesis spectra
of single stellar populations (SSP) libraries by Bruzual \& Charlot (2003) and the code STARLIGHT\footnote{
The STARLIGHT project is supported by the Brazilian agencies CNPq, CAPES and FAPESP and by the France-Brazil CAPES/Cofecub program.}
(Cid-Fernandes et al. 2004, Mateus et al. 2006).  The fitted spectra were later 
subtracted from the observed ones and the emission-line intensities of the residuals were 
compared with the corresponding non-corrected values.
For those objects of our sample with very high H$\beta$ 
equivalent widths (more than 200~{\AA} for WR039 and around 100 {\AA}
for WR038, WR057, WR404, and WR505) the correction at H$\beta$
wavelength is negligible (less than 1~{\AA}). Only in the case of WR266, with EW(H$\beta$)
of 54~{\AA}, we found a correction around 4~{\AA} for EW(H$\beta$). 
For this object, appropriate corrections for each Balmer line were taken into account
according to the SSP fitting in the brightest spaxels.

We calculated the statistical errors of the line fluxes, $\sigma_{l}$, using the
expression $\sigma_{l}$ =
$\sigma_{c}$N$^{1/2}$[1+EW/N$\Delta$]$^{1/2}$ (as in P\'erez-Montero \& D\'\i az 2003)  where
$\sigma_{c}$ represents a standard deviation of the noise in a range centred close to
the measured emission line, N is the number of pixels used in the
measurement of the line flux, EW is the equivalent width of the line,
and $\Delta$ is the wavelength dispersion in \AA/pixel. This
expression takes into account the error in the continuum and the
photon count statistics of the emission line. The error
measurements were performed on the extracted one-dimensional spectra.
In order to minimize errors in the ratios between a
certain emission line and H$\beta$,
we always took first its ratio in relation to the closest hydrogen emission line 
({\em i.e.} H$\alpha$ in the case of [N{\sc ii}])
and then we renormalized using the corresponding theoretical ratio ({\em i.e.}
at the electron temperature derived in the integrated SDSS observations for each
object) from Storey \& Hummer (1995). We checked that the variation of this temperature across the
FoV of the instrument does not introduce errors in the theoretical
ratio larger than those associated with the flux of the emission lines.

\begin{figure*}[h]
\centering
     \includegraphics[width=6cm,clip=]{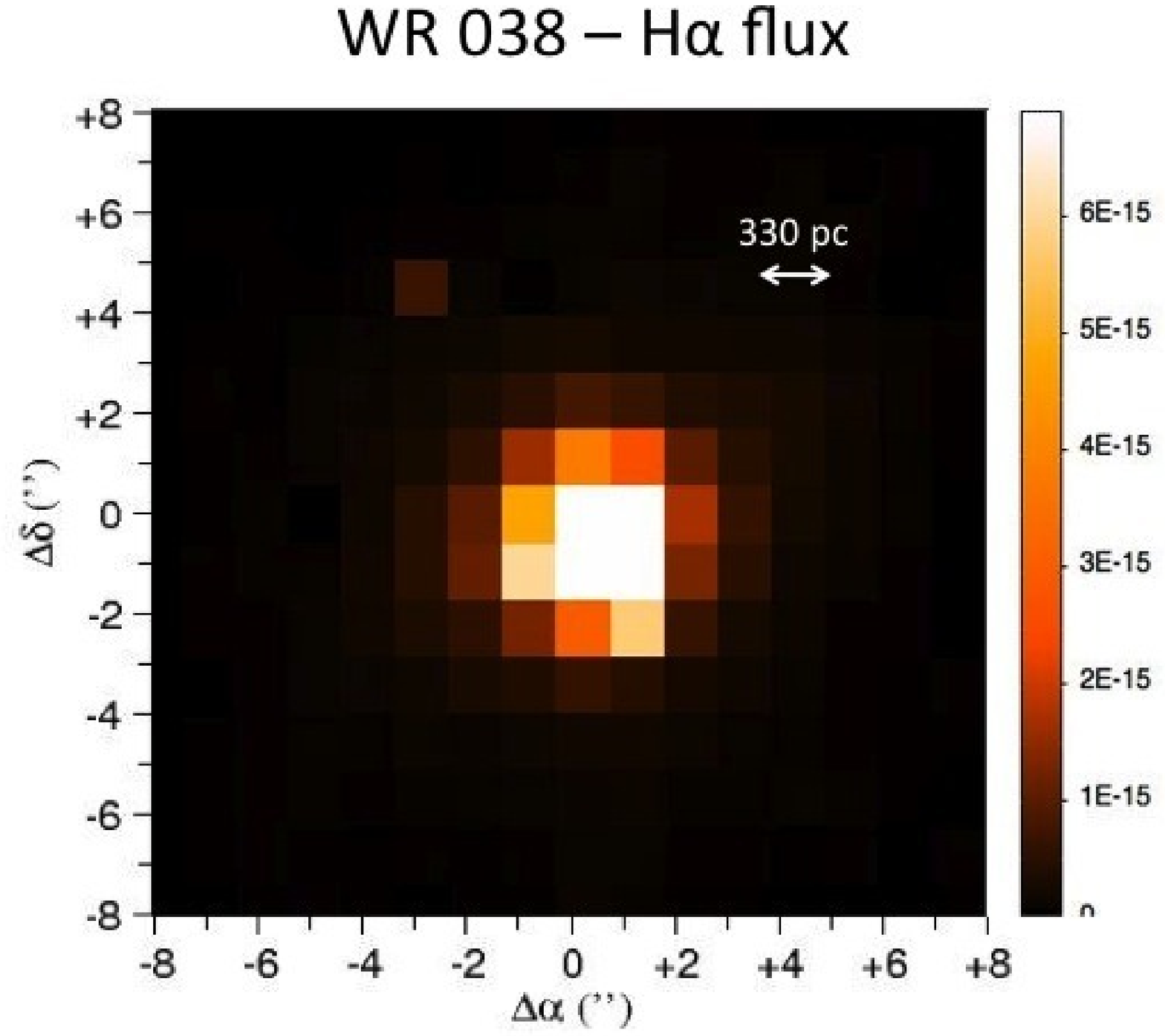}
     \includegraphics[width=6cm,clip=]{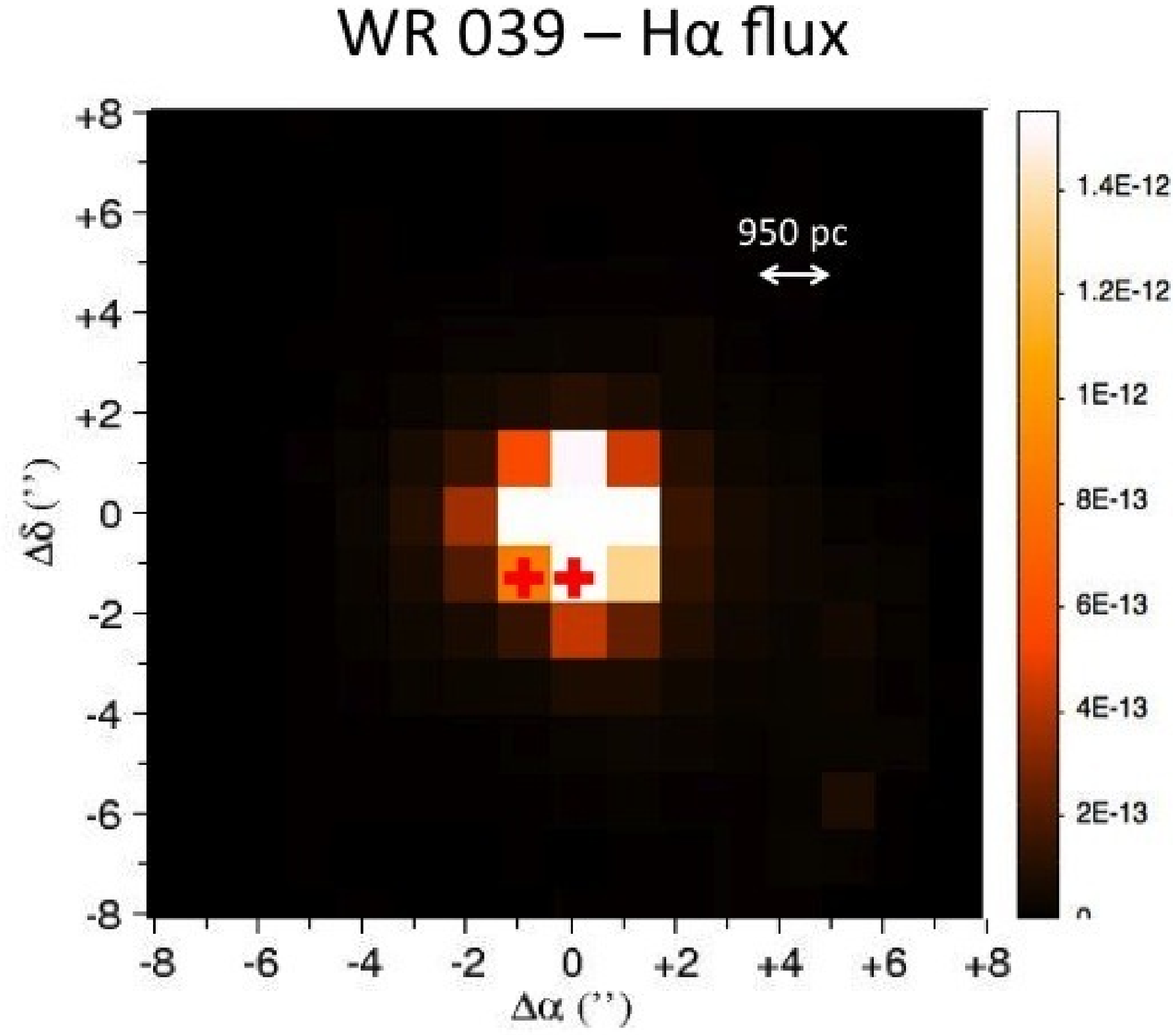}
     \includegraphics[width=6cm,clip=]{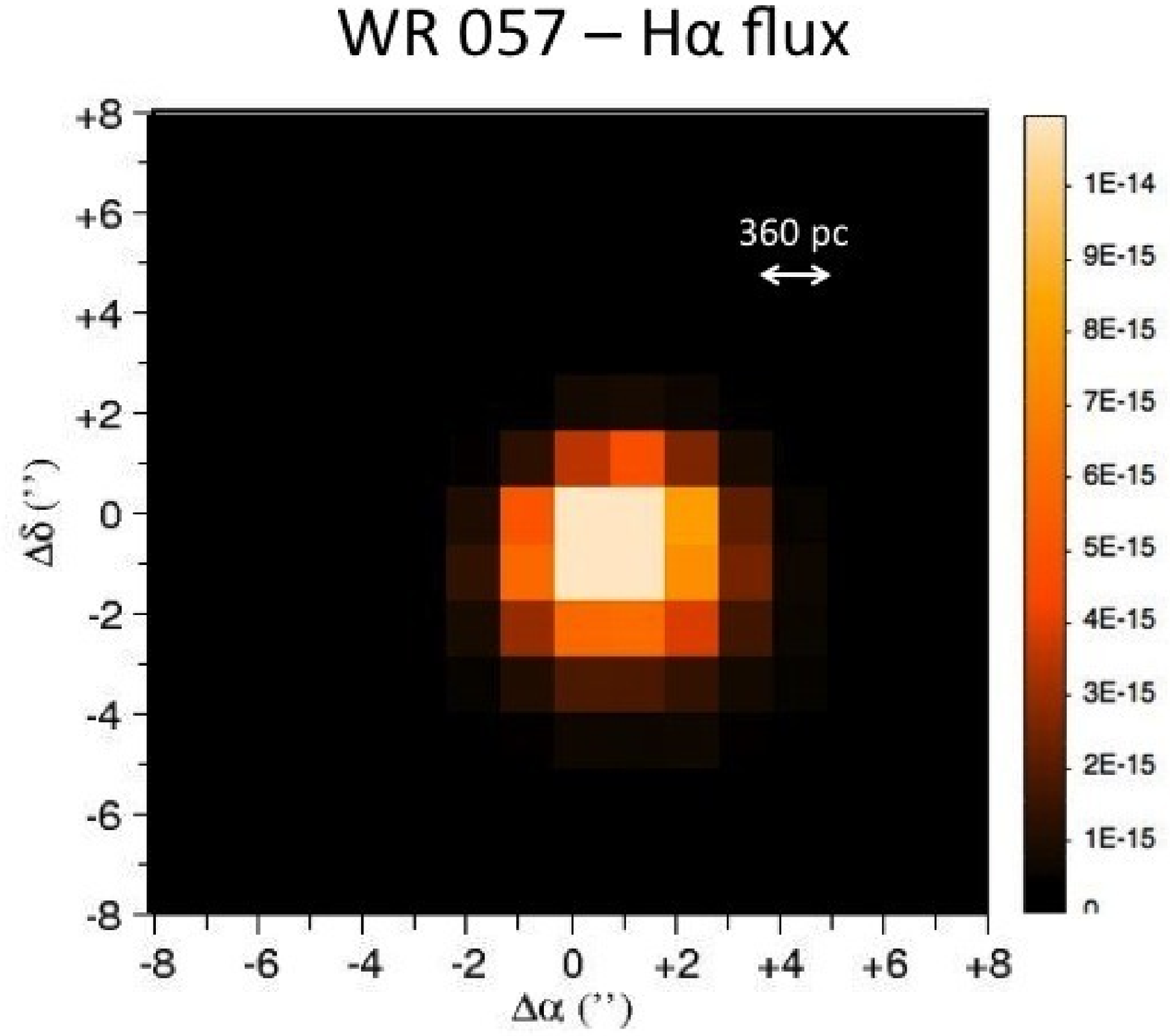}
      \includegraphics[width=6cm,clip=]{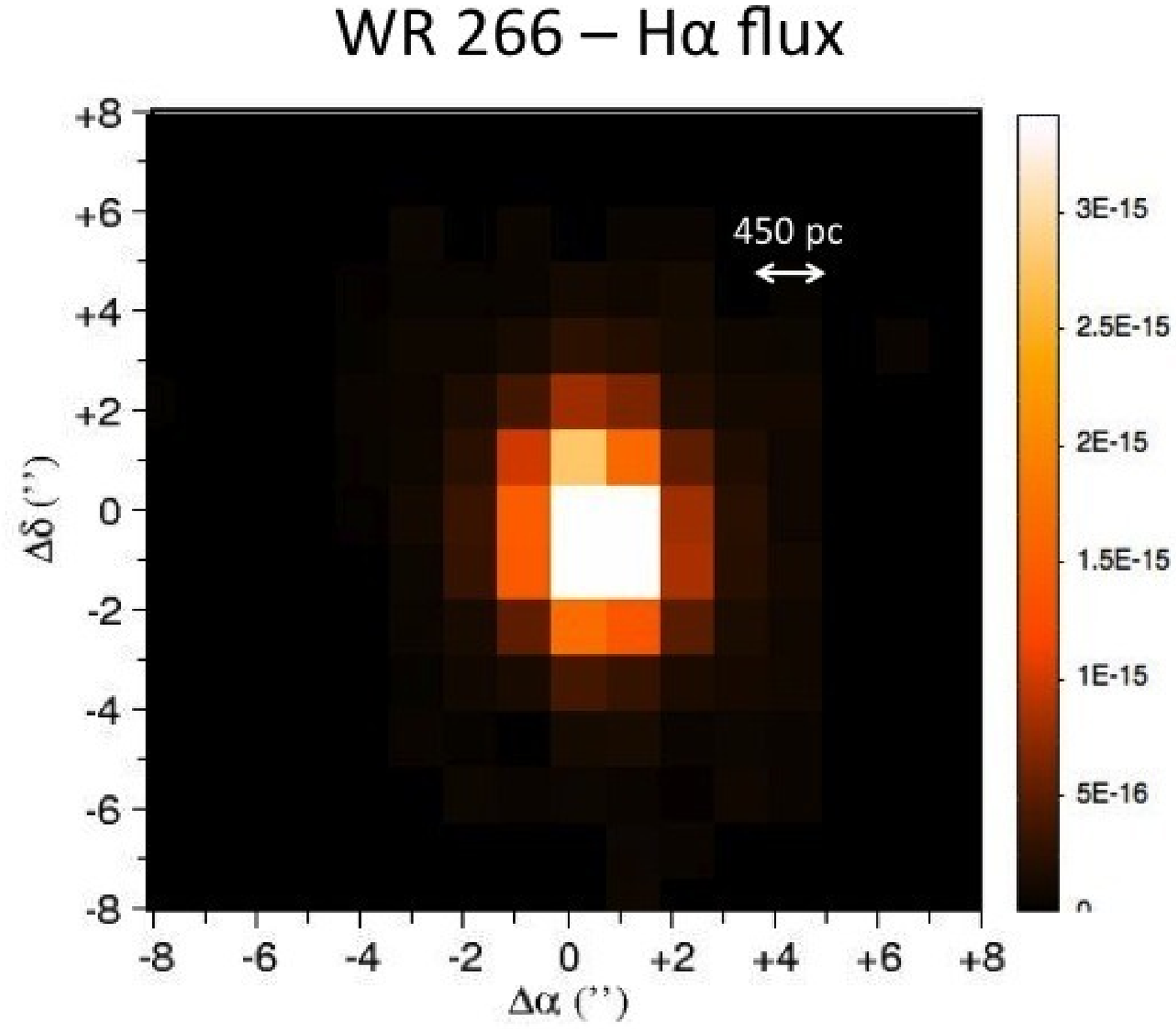}
     \includegraphics[width=6cm,clip=]{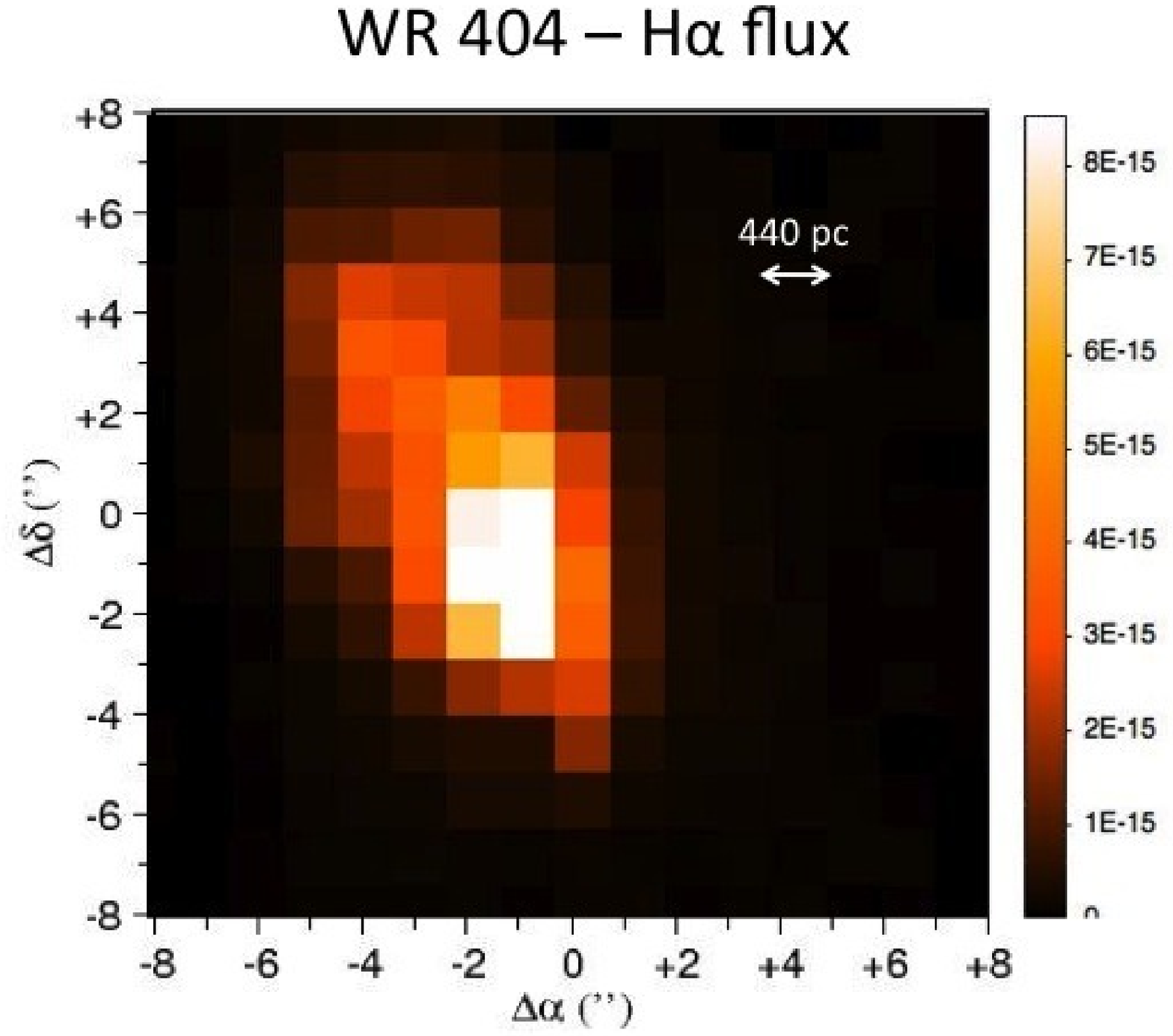}
     \includegraphics[width=6cm,clip=]{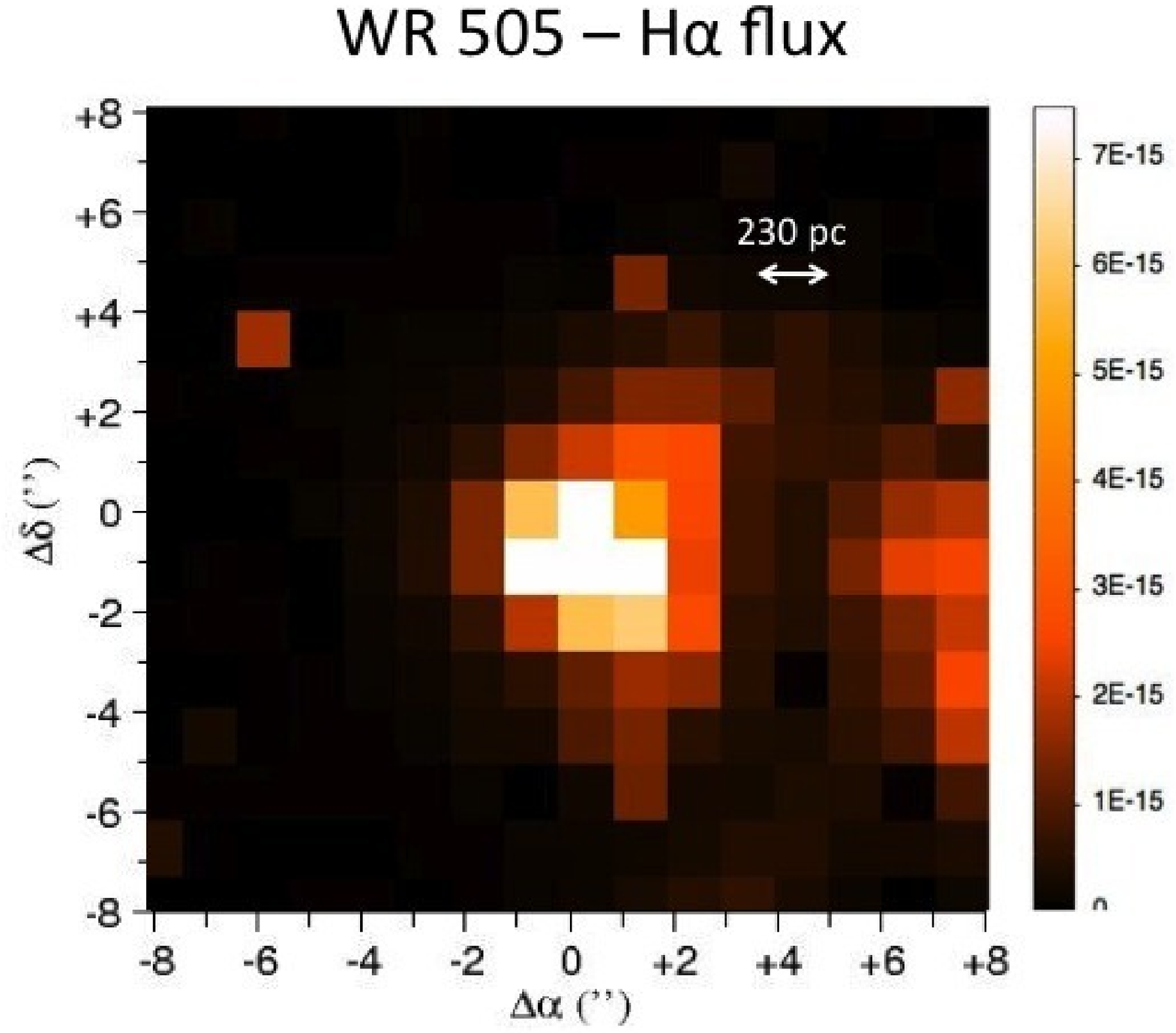}
     
     \caption{Extinction corrected H$\alpha$ maps of the six observed WR galaxies. In all images, each spaxel has 1'' of  resolution. North is up and east to the left. Fluxes are in units of erg/s/cm$^ 2$. The relative size in parsecs, at the adopted distance of each object, is indicated. For WR038, the spaxels where the WR bump was detected from our PMAS data are marked with red crosses.}

    \label{Halfa}
    \end{figure*}

\subsection{Extinction correction and H$\alpha$ maps}

For each fiber spectrum we derived its corresponding reddening
coefficient, c(H$\beta$), using a weighted fit to the values of the Balmer decrement
derived from H$\alpha$/H$\beta$ and H$\gamma$/H$\beta$
as compared to the theoretical values
expected for recombination case B from Storey $\&$ Hummer (1995)
at the electron density and temperature obtained from the
integrated SDSS DR-7 spectra and applying the extinction law given by 
Cardelli et al. (1989) with R$_V$ = 3.1.  In all cases, homogeneous 
low values of the reddening constant 
were derived in agreement with the same values derived from
the analysis of the corresponding SDSS one-dimensional spectra.

The fluxes of the emission lines for each spaxel were corrected for extinction
using its corresponding c(H$\beta$) value. 
H$\alpha$ emission line maps (continuum subtracted and extinction corrected) 
are shown in Fig.~\ref{Halfa}.  As can be seen the observed FoV 
encompasses the whole optical extent of
our galaxies, which are mostly very compact, with the exception of WR404,
which presents a cometary aspect, with a low brightness tail towards the NE
direction, and WR 505, which presents several knots of star formation other
than the brightest one at west of the FoV.

\subsection{Oxygen and nitrogen chemical abundances}

\begin{figure*}
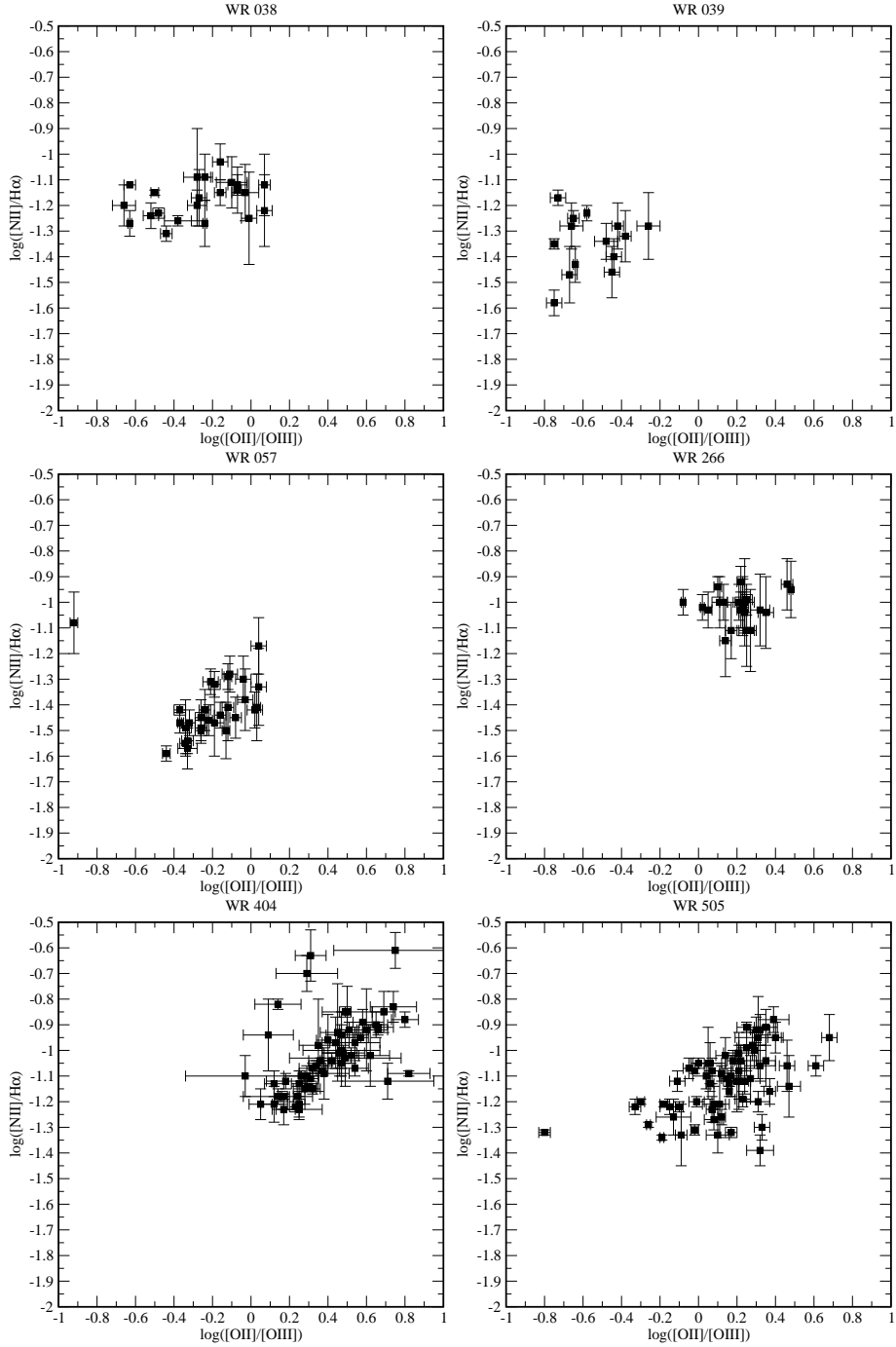

\centering
     \includegraphics[width=6cm,clip=]{wr038_O2O3_N2.eps}
     \includegraphics[width=6cm,clip=]{wr039_O2O3_N2.eps}
     \includegraphics[width=6cm,clip=]{wr057_O2O3_N2.eps}
      \includegraphics[width=6cm,clip=]{wr266_O2O3_N2.eps}
     \includegraphics[width=6cm,clip=]{wr404_O2O3_N2.eps}
     \includegraphics[width=6cm,clip=]{wr505_O2O3_N2.eps}
     
     \caption{Relation between the \oii/\oiii and \nii/H$\alpha$ emission line ratios
for those spaxels with sufficient S/N in the four involved lines of the
six observed galaxies. }

    \label{exc}
    \end{figure*}

The chemical abundances of oxygen and nitrogen were studied in a
representative sample of the observed spaxels in the six WR galaxies
using different methods as explained below.

The most accurate method to derive oxygen abundances in emission-line
objects, as those in our sample, is the so-called direct method which
depends on the relative intensity of both nebular oxygen emission lines to
a hydrogen recombination emission-line, such as H$\beta$ ({\em i.e.} 
\oii/{\hbeta} for O$^+$/H$^+$ and \oiii/{\hbeta} for O$^{2+}$/H$^+$), and the previous
determination of the electron temperature, via the quotient between 
auroral-to-nebular emission lines, such as \oiii~4363 {\AA}
and \oiii~5007 \AA.  
See, for instance, P\'erez-Montero et al. (2011) or
Kehrig et al. (2013) for additional details of
this procedure and how to calculate the low-excitation electron temperature
to derive low-excitation ionic abundances. This method was applied to the SDSS DR7 spectra of the brightest
regions of the six observed galaxies leading to values of the total oxygen
abundances compatible with the low metallicity regime. 
Owing to the spectral coverage in the SDSS spectra (3800-9100~\AA), the
\oii~3727 {\AA} emission-line was not detected in WR038, WR057, and WR505, 
and the \oii~7319,7330 {\AA}
were used instead, as described in
Kniazev et al. (2002). The total abundances derived in the SDSS spectra
using this method are listed in Table \ref{abs}.

Regarding the PMAS IFS observations, the direct method was
applied in a representative number of spaxels only in
WR039, where the \oiii~4363 {\AA} emission line was detected
with acceptable S/N ($>$ 2.5) . In the other galaxies of our sample, the direct method could only be applied
in the brightest spaxels in WR038, WR057, and WR505 and in none
of them in WR266 and WR404.

Therefore, the spatial analysis of the chemical
properties in these galaxies was done by means of strong-line methods.
We first resorted to the N2 parameter ({\em e.g.} Denicol\'o et al. 2002),
defined as the ratio between \nii~6584 {\AA} and H$\alpha$. This parameter has
the advantage that it does not depend  on reddening nor flux calibration uncertainties
and is linearly well correlated with oxygen abundance up to solar metallicities.
On the contrary it has an important drawback when it is used for extended objects,
as those studied in this work by means of IFS, as it also varies as a function
of the excitation conditions (P\'erez-Montero \& D\'\i az 2005). We verified in what 
objects this method could be applied 
to derive reliably the  spatial distribution of the oxygen abundance, by plotting in
Figure \ref{exc} the relation between the N2 parameter and the \oii/\oiii~ratio,
which traces the nebular excitation. As can be seen, in the most extended
objects of our sample, WR404 and WR505, there is a clear correlation between these
two emission-line ratios. In the other objects there is no
clear  relation between them, with the possible exception of WR057,
but in this object the spatial variation of N2 is lower than the observational errors.
Hence, in the case of WR404 and WR505 galaxies,
we used the strong-line parameter O3N2, firstly introduced as a metallicity 
calibrator by Alloin et al. (1979) and which is defined as the emission-line ratio between
 \oiii~5007 {\AA} and \nii~6584 {\AA}. According to several authors, such as Pettini \&
Pagel (2004) or P\'erez-Montero \& Contini (2009), this parameter is not valid for
very low metallicity objects (12+log(O/H) $<$ 8.0) but, on the contrary, its
dependence on excitation is much lower than in the case of N2. According to
the values derived from the SDSS spectra for WR404 and WR505, their mean
oxygen abundances are higher than the lower limit for O3N2, so this parameter
was used for these two objects instead.

For the sake of consistency between the three employed methods (direct
method in the case of WR039, N2 parameter for WR038, WR057, and WR266,
and O3N2 for WR404 and WR505) we used the calibrations presented in
P\'erez-Montero \& Contini (2009) for N2 and O3N2, which are consistent with the
direct method.
The resulting oxygen abundance maps of the six observed galaxies are
plotted in Figure \ref{oh} along with the histogram distributions of the
abundances in those spaxels with enough S/N in all the involved
emission lines (S/N $>$ 2.5).

In the case of the nitrogen-to-oxygen ratio (N/O) the direct method can also
be used to derive the N$^+$/H$^+$ ratio and then deriving N/O using
the approximation N$^+$/O$^+$ $\approx$ N/O [as before see further details
in P\'erez-Montero et al. (2011) or Kehrig et al. (2013)], but this method could only
be used in a representative number of spaxels in WR039. For the other five
galaxies we resorted to the N2O2 parameter, defined as the ratio of \nii~6584~{\AA}
and \oii~3727 {\AA}. This ratio has the advantage that it has a monotonic linear
relation with N/O and, contrary to N2, it does not have any dependence on
excitation as it only depends on low-excitation emission-lines. As in the case
of oxygen abundances, we used the empirical calibration of N2O2 with N/O
from P\'erez-Montero \& Contini (2009), which is consistent with the direct
method derivations of this chemical abundance ratio.
The resulting N/O maps of the six observed galaxies are
plotted in Figure \ref{no} along with the histogram distributions of the
abundances in those spaxels with enough S/N in all the involved
emission lines.

\begin{figure*}
\centering
     \includegraphics[width=7cm,clip=]{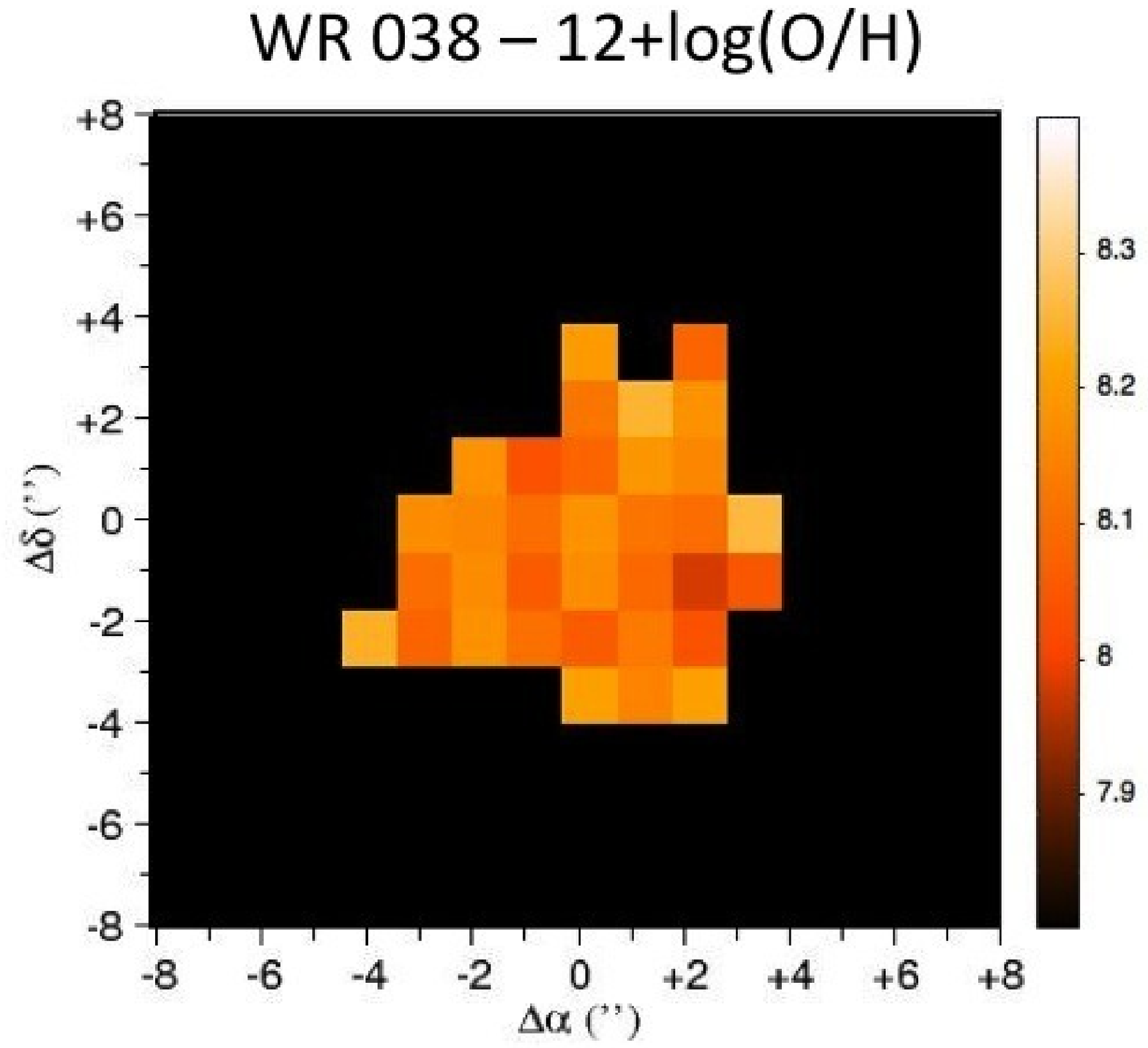}
     \includegraphics[width=6cm,clip=]{wr038_OHe_hist.eps}
     \includegraphics[width=7cm,clip=]{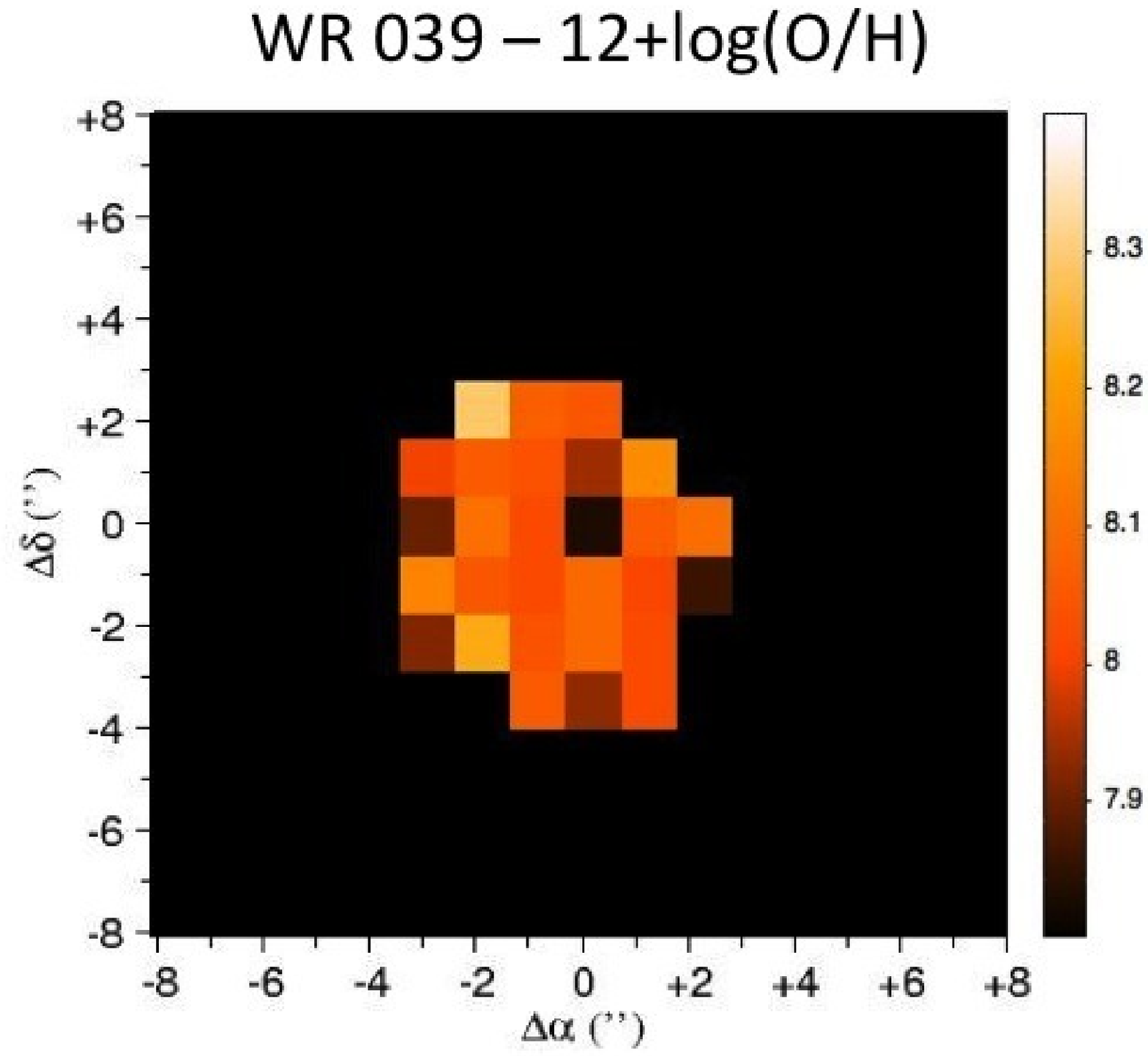}
      \includegraphics[width=6cm,clip=]{wr039_OHd_hist.eps}
     \includegraphics[width=7cm,clip=]{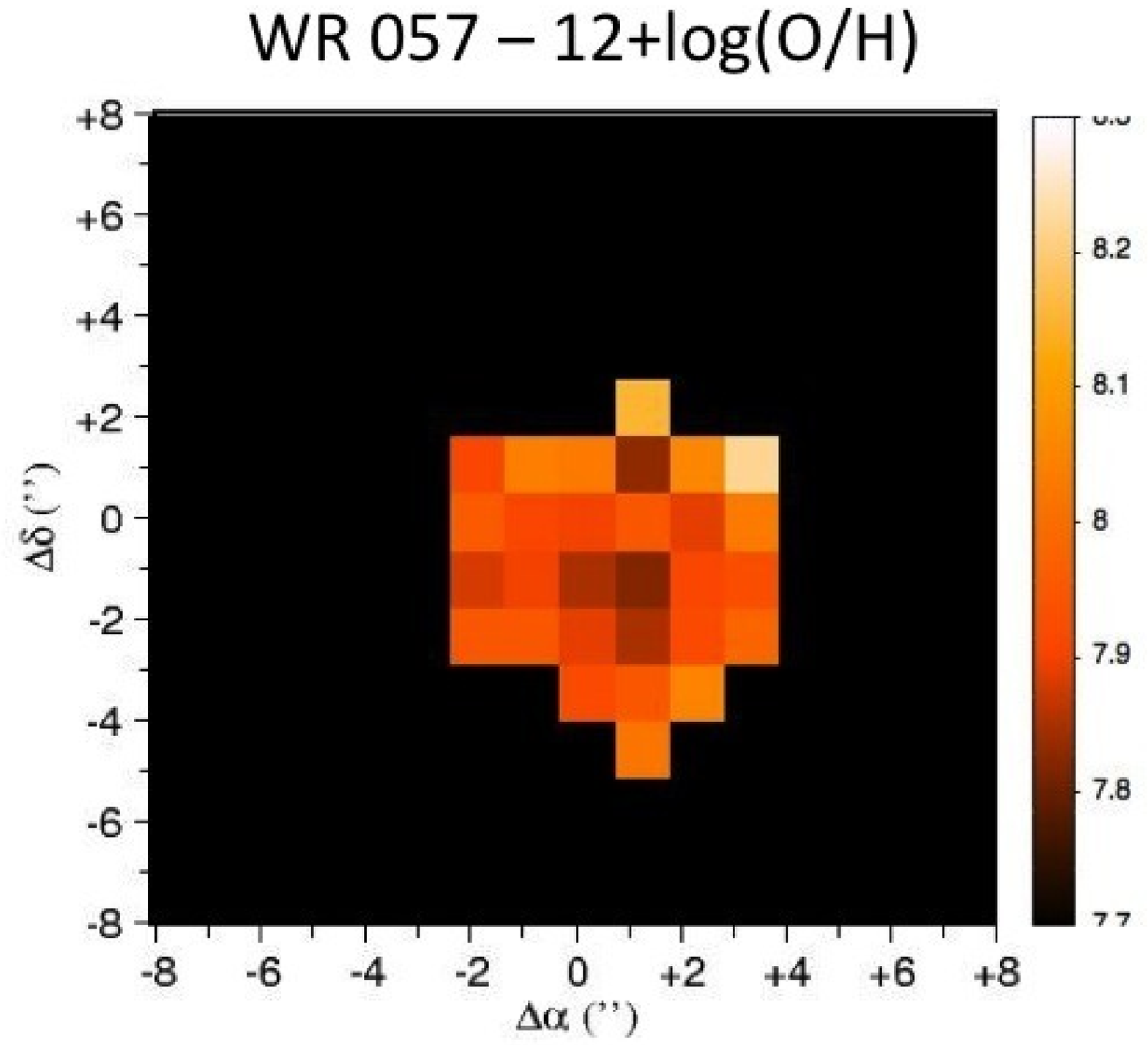}
     \includegraphics[width=6cm,clip=]{wr057_OHe_hist.eps}
     
     \caption{Oxygen abundance maps derived as described in the text and 
histogram distributions in linear scale for WR038, WR039, and WR057. 
   In all images, each spaxel has 1'' resolution, north is up and east is left. 
The bars in the histogram represent spaxels with detected WR emission (red),
spaxels in the same positions as the SDSS pointing (black),  spaxels adjacent
to this position (brown), and the other spaxels (grey).}

    \label{oh}
    \end{figure*}

\begin{figure*}
\centering
\setcounter{figure}{2}
     \includegraphics[width=7cm,clip=]{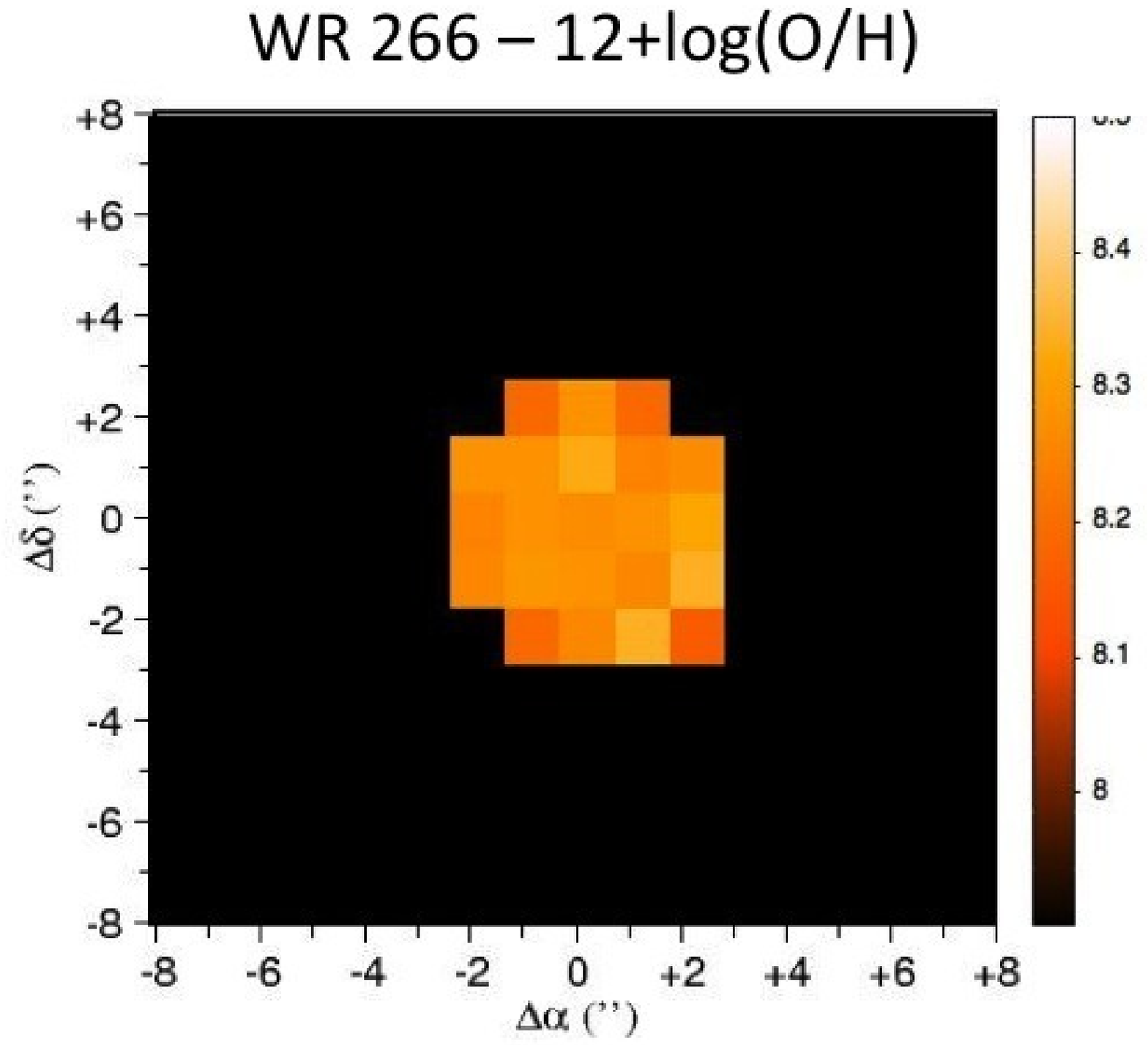}
     \includegraphics[width=6cm,clip=]{wr266_OHe_hist.eps}
     \includegraphics[width=7cm,clip=]{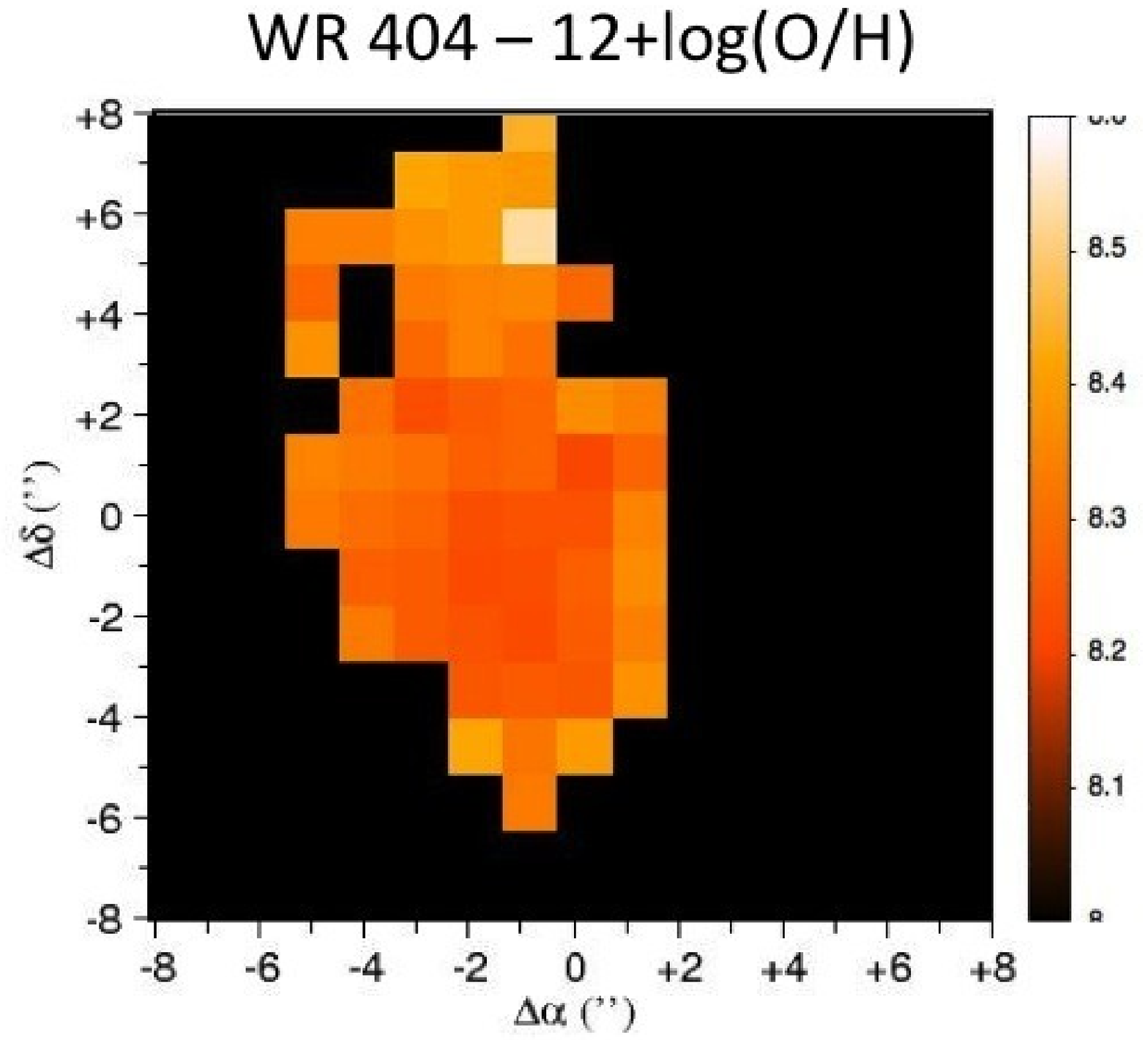}
      \includegraphics[width=6cm,clip=]{wr404_OHe_hist.eps}
     \includegraphics[width=7cm,clip=]{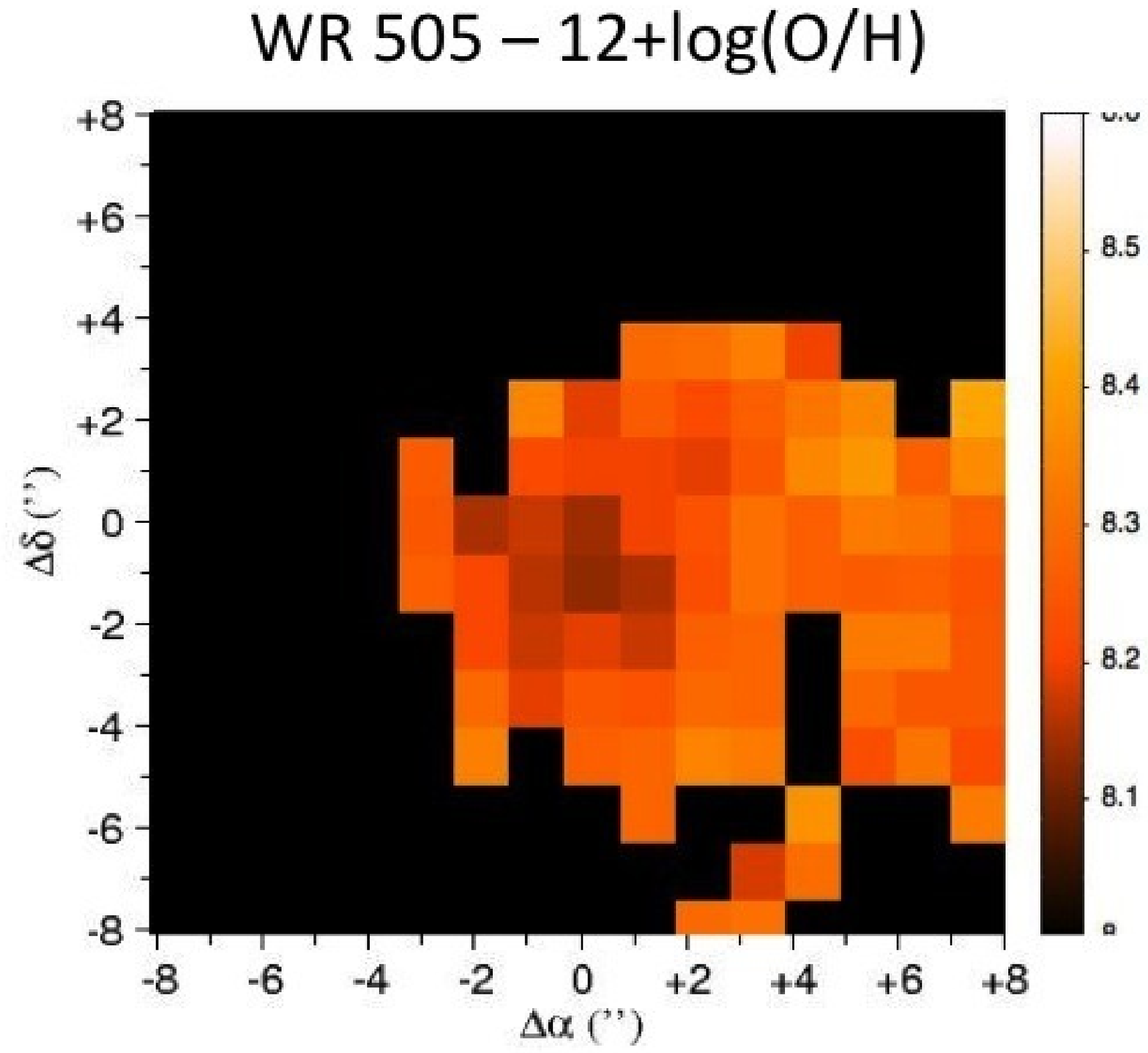}
     \includegraphics[width=6cm,clip=]{wr505_OHe_hist.eps}
     
     \caption{ (cont.) Same figure for WR266, WR404, and WR505}

    \label{OHe}
    \end{figure*}

\begin{figure*}
\centering
     \includegraphics[width=7cm,clip=]{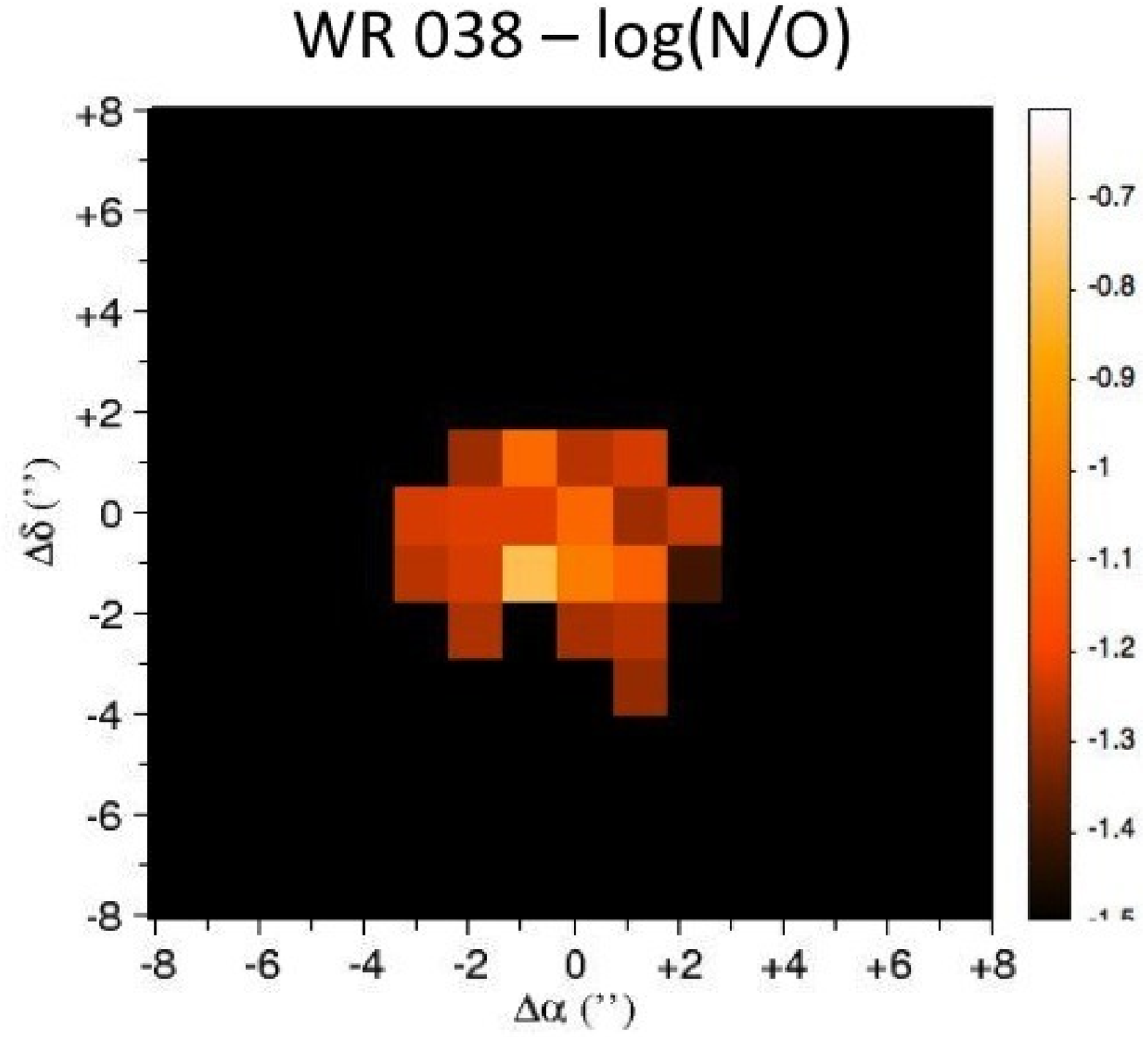}
     \includegraphics[width=6cm,clip=]{wr038_NOe_hist.eps}
     \includegraphics[width=7cm,clip=]{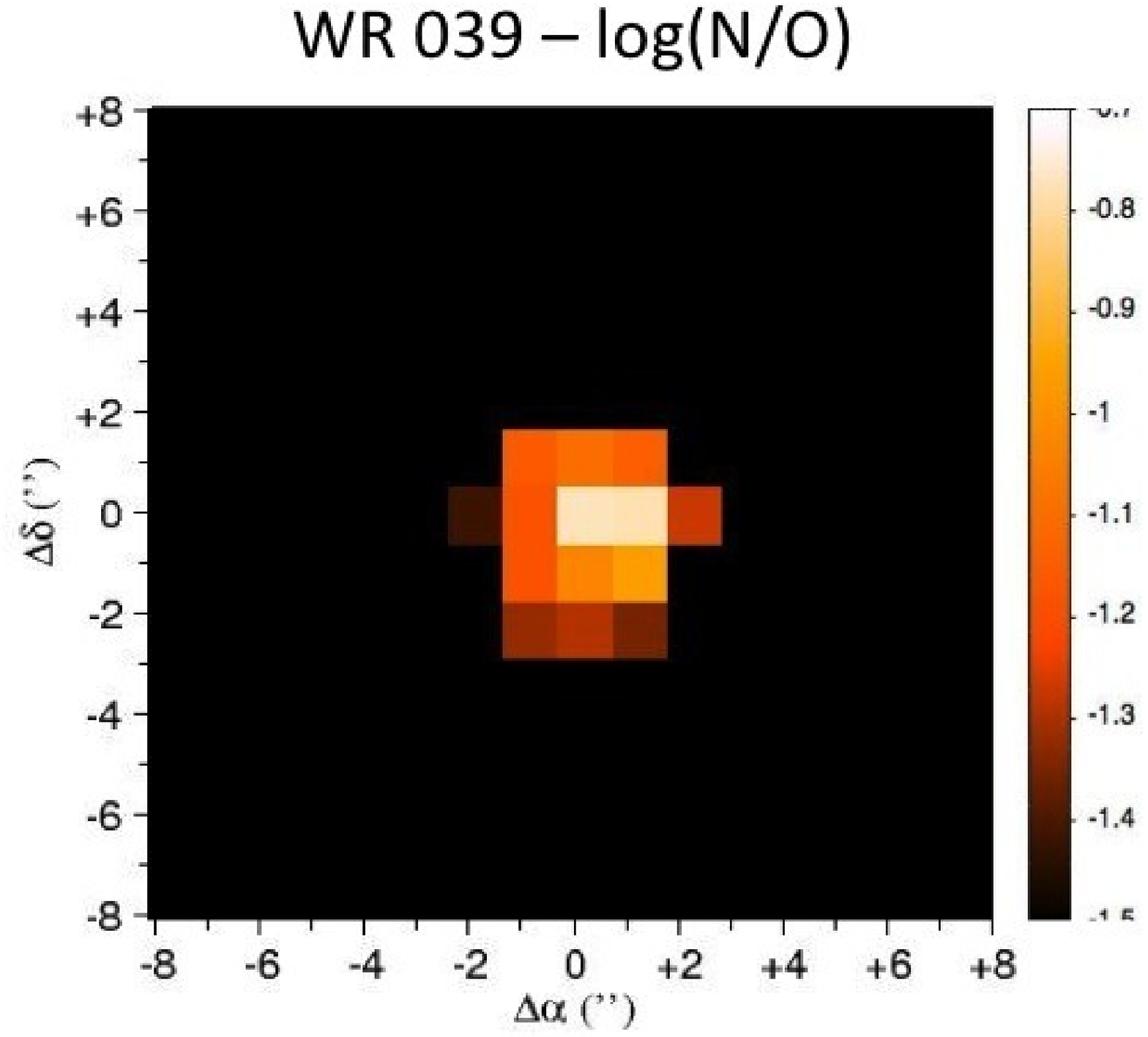}
      \includegraphics[width=6cm,clip=]{wr039_NOd_hist.eps}
     \includegraphics[width=7cm,clip=]{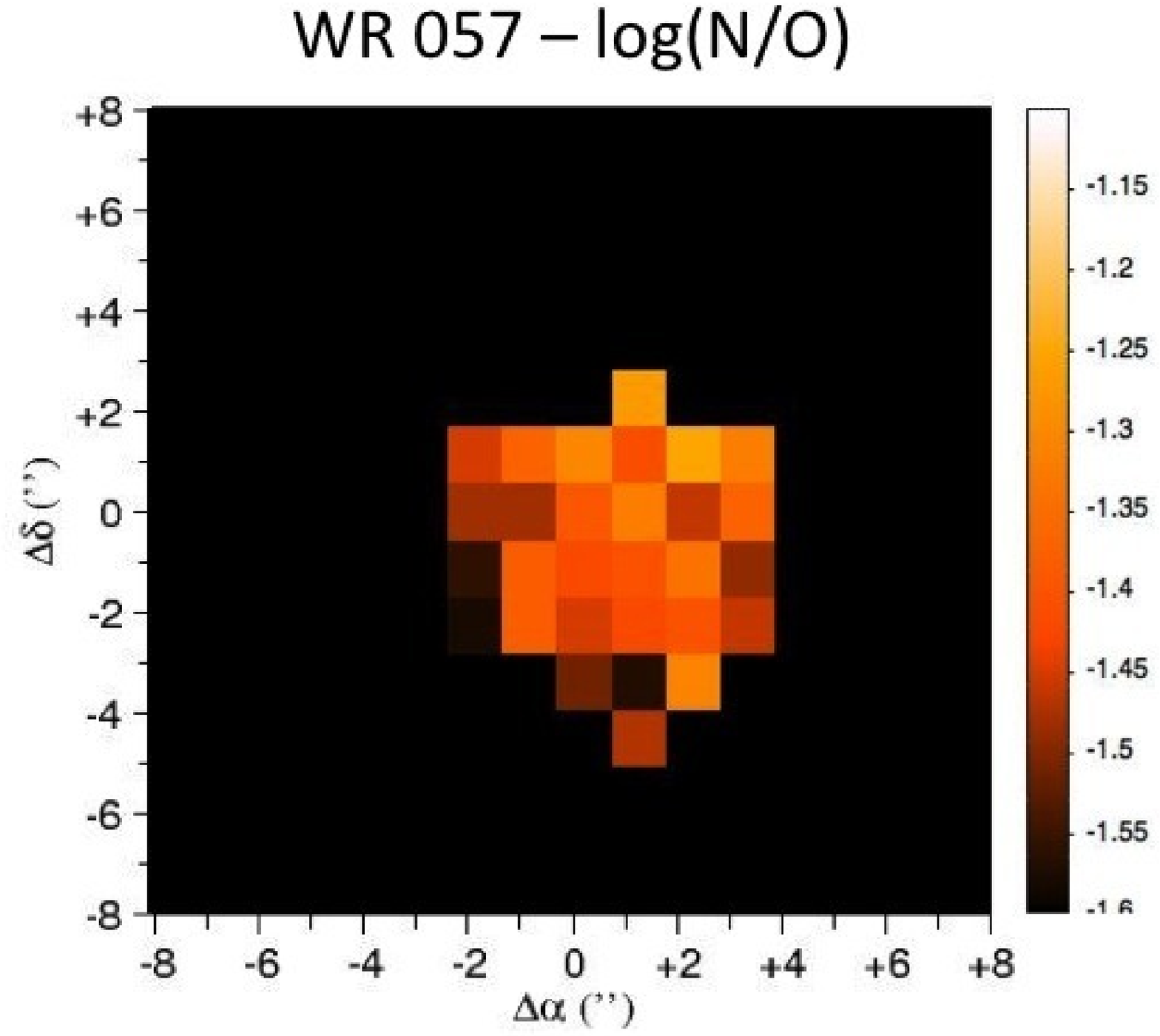}
     \includegraphics[width=6cm,clip=]{wr057_NOe_hist.eps}
     
     \caption{N/O ratio maps derived as described
in the text and histogram distributions in linear scale for WR038, WR039, and WR057.
In all images, each spaxel has 1''  resolution, north is up and east is left. The colors
in the bars have the same meaning as in Figure 3.}

    \label{no}
    \end{figure*}

\begin{figure*}
\centering
\setcounter{figure}{3}
     \includegraphics[width=7cm,clip=]{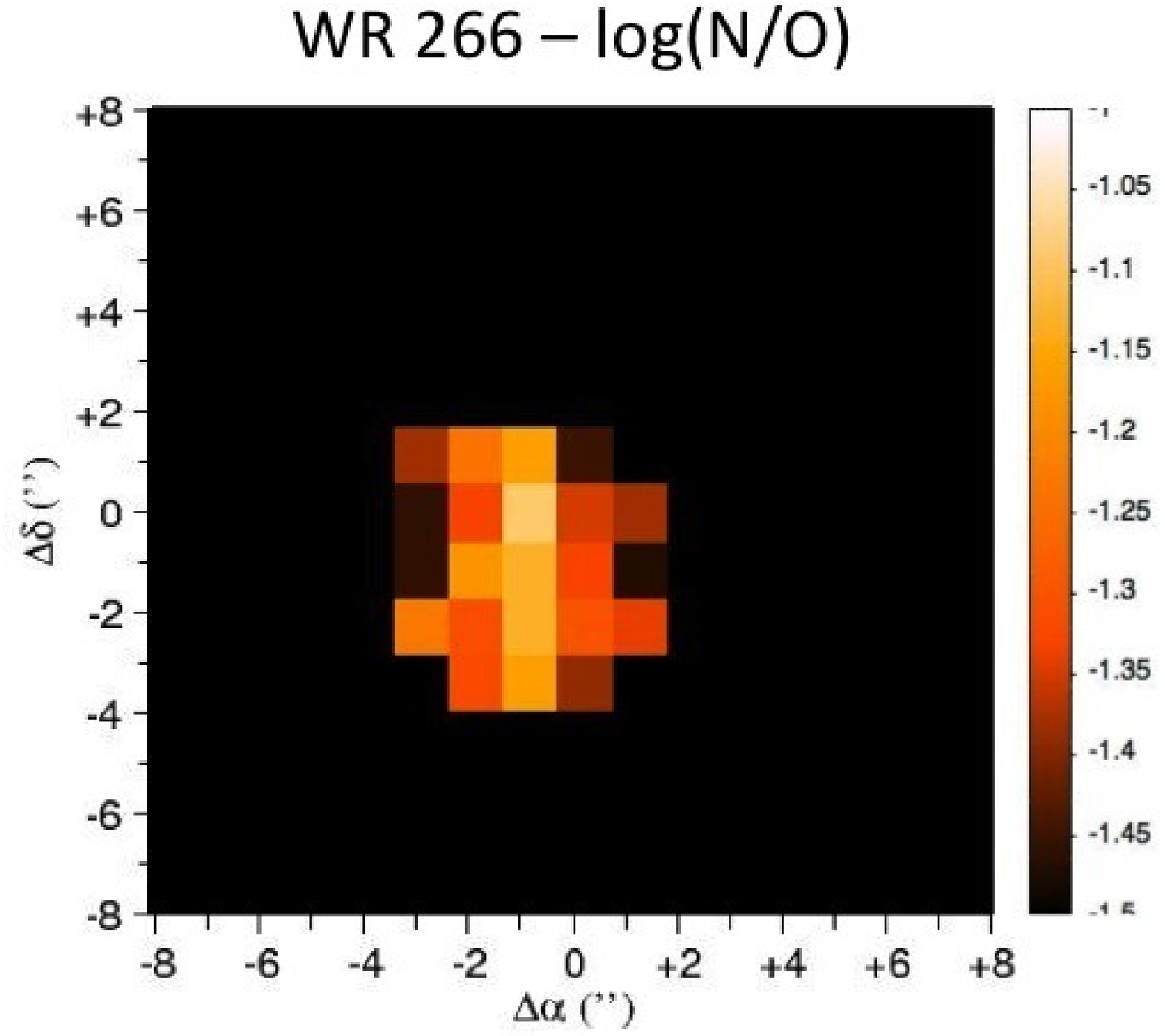}
     \includegraphics[width=6cm,clip=]{wr266_NOe_hist.eps}
     \includegraphics[width=7cm,clip=]{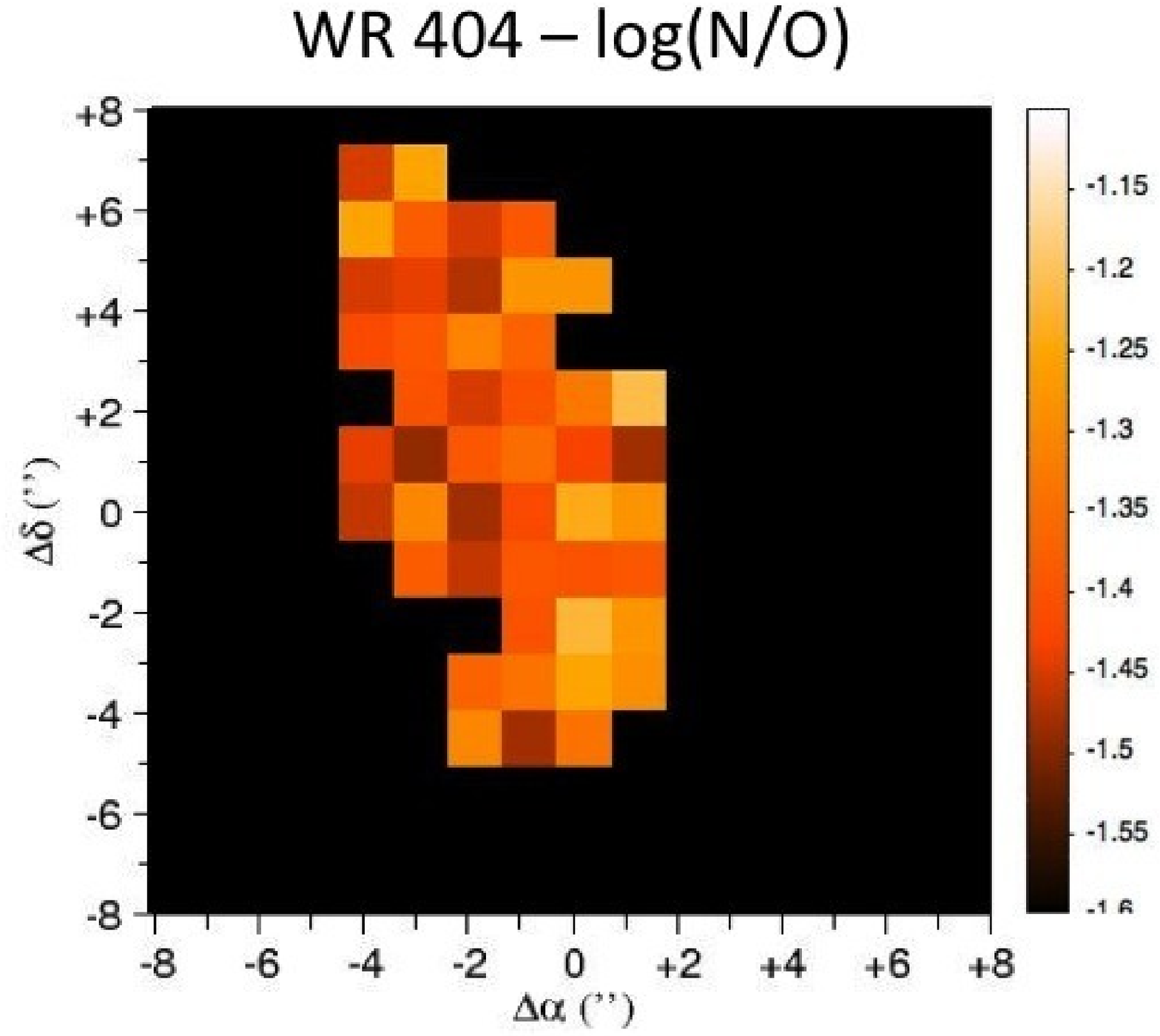}
      \includegraphics[width=6cm,clip=]{wr404_NOe_hist.eps}
     \includegraphics[width=7cm,clip=]{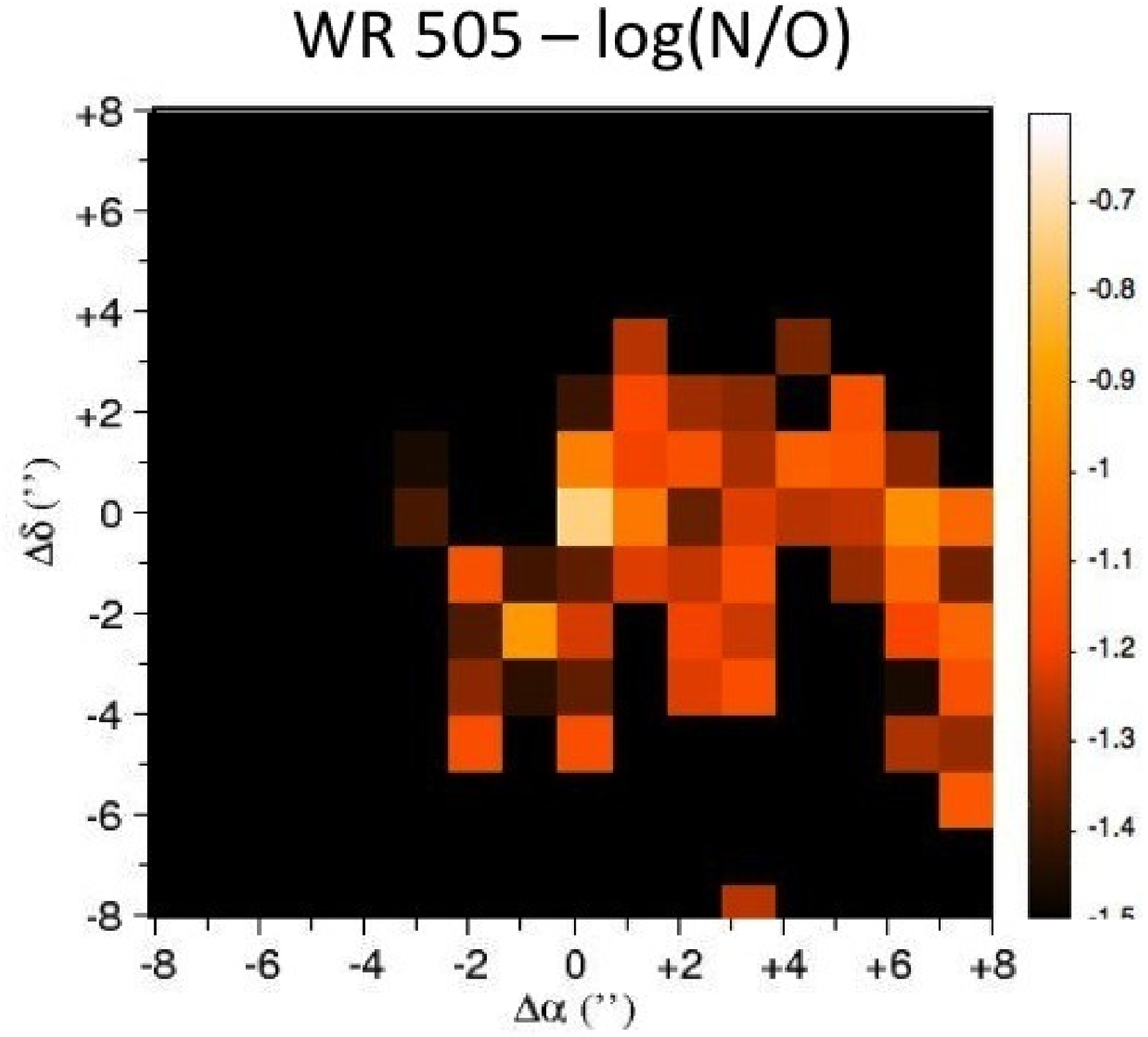}
     \includegraphics[width=6cm,clip=]{wr505_NOe_hist.eps}
     
     \caption{(cont.)  Same figure for WR266, WR404, and WR505}
  
    \label{no}
    \end{figure*}

\begin{table*}

\centering
\caption{Total oxygen abundances and N/O derived
for the six studied WR galaxies using different methods as described in the
text. The confidence level from the Lilliefors test for those 
spaxels where all involved emission-lines were measured is the p-value. 
Column (3) lists the mean value and standard deviation (sigma) from the Gaussian 
fit as long as p-value $>$ 0.05; otherwise the mean value and sigma of the data
distribution are shown. Column (4) and (5) show the O/H and N/O, and their 
corresponding errors derived from the brightest spaxel and from the SDSS spectrum 
for each galaxy, respectively. The number in parenthesis indicates the method used
to derive the chemical abundances: (1) from direct method with \oii~3727 {\AA},
(2) from direct method with \oii~7319,7330 {\AA}, (3) from N2, (4) from O3N2, and (5) from N2O2 .}
\begin{tabular} {lcccc}
\hline
  & p-value & mean $\pm$ st. deviation  &  brightest  & SDSS  \\
 \hline
12+log(O/H) \\
WR 038 & 0.51 & 8.14 $\pm$ 0.06 (3) & 8.09 $\pm$ 0.07 (1) & 8.16 $\pm$ 0.11 (2) \\
WR 039 & 0.07 & 8.06 $\pm$ 0.09 (1) & 7.96 $\pm$ 0.05 (1) & 8.13 $\pm$ 0.05 (1) \\
WR 057 & 0.10 & 7.94 $\pm$ 0.06 (3) & 8.14 $\pm$ 0.10 (1) & 8.06 $\pm$ 0.06 (2) \\
WR 266 & 0.23 & 8.23 $\pm$ 0.05 (3) & 8.26 $\pm$ 0.30 (1) & 8.18 $\pm$ 0.14 (1) \\
WR 404 & 0.02 & 8.26 $\pm$ 0.02 (4) & 8.23 $\pm$ 0.30 (4) & 8.23 $\pm$ 0.05 (1) \\
WR 505 & 0.30 & 8.31 $\pm$ 0.06 (4) & 8.16 $\pm$ 0.16 (4) & 8.09 $\pm$ 0.08 (2) \\

\hline
log(N/O) \\ 
WR 038 & 0.01 & -1.14 $\pm$ 0.14 (5) & -0.87 $\pm$ 0.27 (1) & -1.06 $\pm$ 0.24 (2) \\
WR 039 & 0.04 & -1.09 $\pm$ 0.09 (1) & -0.80 $\pm$ 0.20 (1) & -1.17 $\pm$ 0.14 (1) \\
WR 057 & 0.39 & -1.42 $\pm$ 0.07 (5) & -1.48 $\pm$ 0.26 (1) & -1.32 $\pm$ 0.15 (2) \\
WR 266 & 0.04 & -1.29 $\pm$ 0.10 (5) & -1.13 $\pm$ 0.30 (1) & -1.18 $\pm$ 0.24 (1) \\
WR 404 & 0.00 & -1.37 $\pm$ 0.07 (5) & -1.39 $\pm$ 0.30 (5) & -1.40 $\pm$ 0.06 (1) \\
WR 505 & 0.00 & -1.21 $\pm$ 0.15 (5) & -1.15 $\pm$ 0.38 (5) & -1.27 $\pm$ 0.16 (2) \\
\hline
\end{tabular}
\label{abs}
\end{table*}

\subsection{Spatial chemical homogeneity}

To study to what extent the chemical content of 
the gas can be considered as homogeneous, and to give the statistical significance 
of the O/H and N/O distributions, we used the procedure presented in
P\'erez-Montero et al. (2011) and refined in Kehrig et al. (2013).
This method is based on the assumption that a certain
property can be considered as spatially homogeneous across 
the observed FoV if
two conditions are satisfied: for the corresponding dataset (i) 
the null hypothesis
({\em i.e.} the data come from a normally distributed population) of the
Lilliefors test (Lilliefors, 1967) cannot be rejected at the 5$\%$
significance level, and (ii) the observed variations of the data distribution around the single
mean value can be explained by random errors;
{\em i.e.} the corresponding Gaussian sigma  should be
lower or of the order of the typical uncertainty of the considered
property; we take as typical uncertainty the square root of the weighted sample variance.
In Table~\ref{abs} we show the results from our statistical analysis
for both total O/H and N/O.

The Lilliefors test for each of the distributions was performed on
the linear values of the chemical abundances. 
The corresponding confidence levels (p-values)
are listed in Table \ref{abs} with
the resulting means and Gaussian sigma in case that the p-value is higher
than 0.05. Otherwise, the means and standard deviations are those of
the distributions. In those cases where the p-value is larger
that 0.05, the second condition imposed to consider
a homogeneous distribution is also satisfied in all cases.  
In all distributions where a strong-line method
was used to derive both O/H and N/O, the sigma of the Gaussian 
is much lower than the intrinsic uncertainty
associated with these methods ($\sim$ 0.3~dex). For WR039, where
the direct method was used the weighted sigma for both oxygen abundance 
and N/O is 0.2~dex, which also is larger than the usual sigmas in the
Gaussian fittings.

\begin{figure*}
\centering
     \includegraphics[width=7cm,clip=]{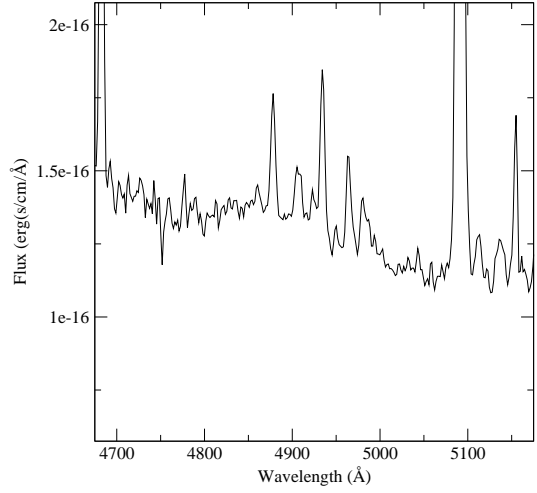}
     \caption{Optical spectrum taken with PMAS coadded in the two spaxels in
WR038 where the Wolf-Rayet blue bump is detected.}

    \label{wrbb}
    \end{figure*}

\begin{figure*}
\centering
     \includegraphics[width=6.5cm,clip=]{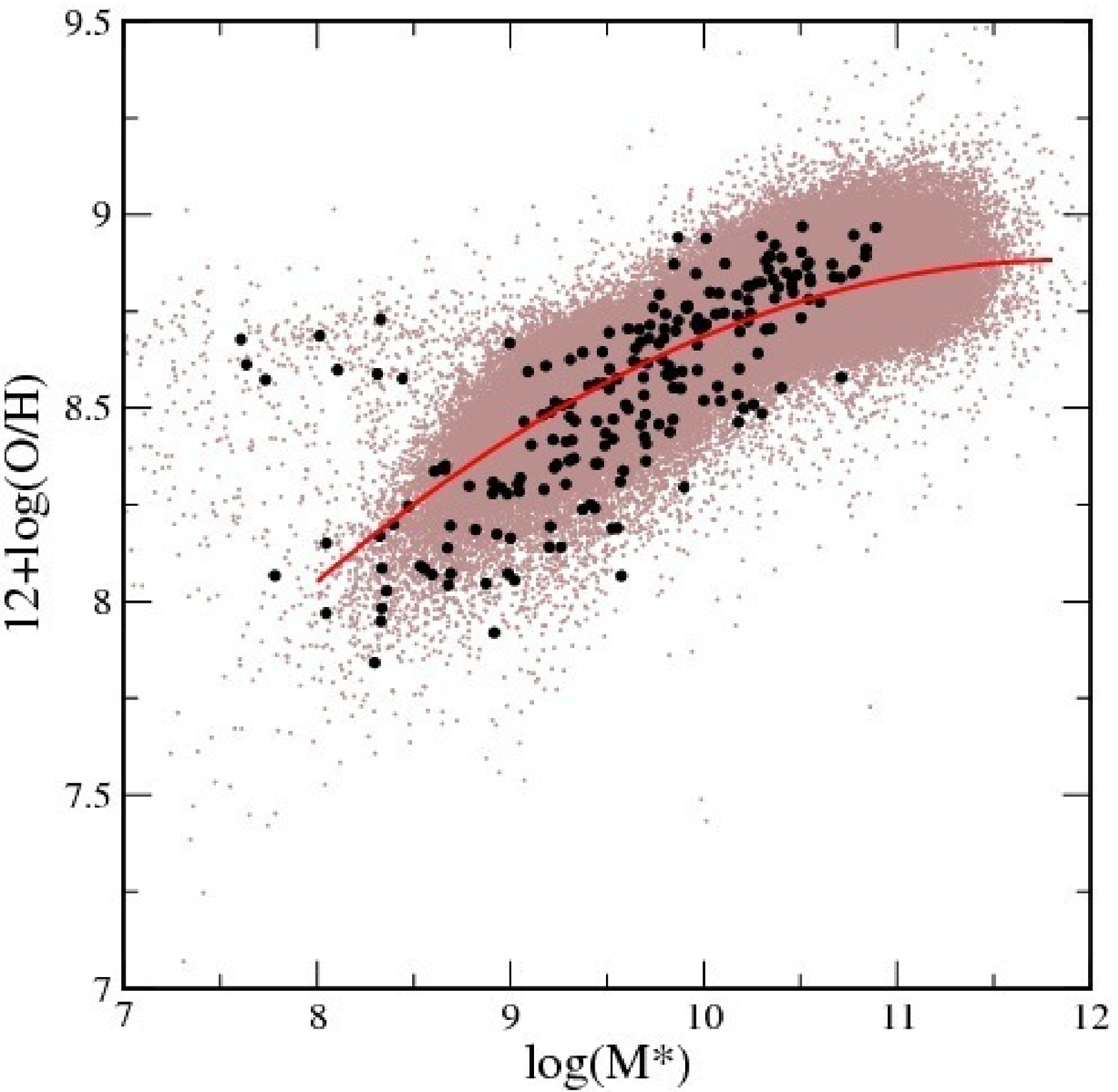}
     \includegraphics[width=6.5cm,clip=]{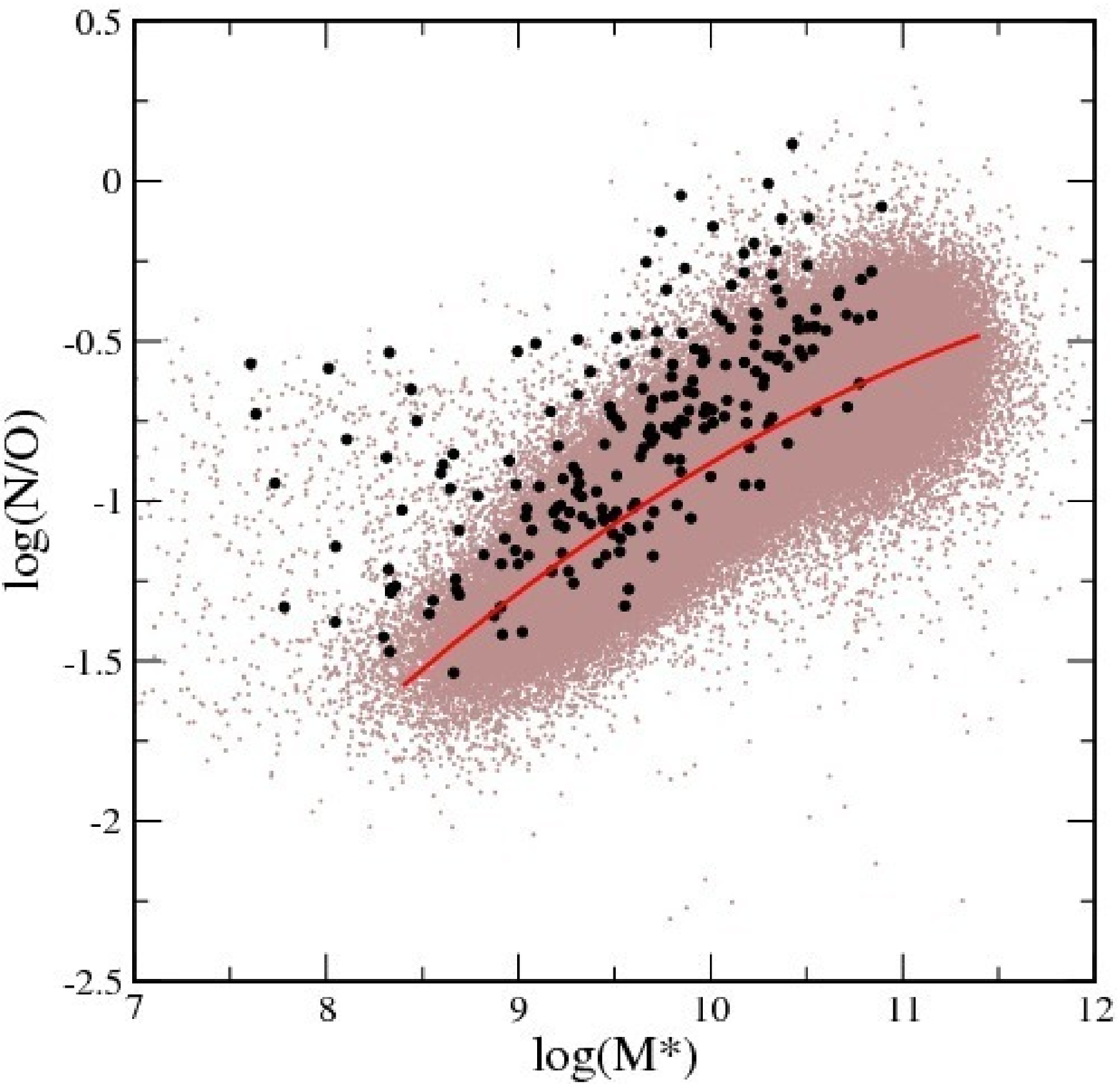}
     
     \caption{MZR (left) and MNOR (right) relations
for the star-forming galaxies selected from the SDSS DR7. The solid red
lines, calculated by P\'erez-Montero et al. (2013), are quadratic fits to the
medians in stellar mass bins of 0.2 dex. The black points are the galaxies
matching the catalog of WR galaxies  by Brinchmann et al. (2008). }

    \label{SDSS_WR}
    \end{figure*}

\section{Discussion: are WR stars able to pollute the ISM of galaxies?}

The use of IFS is fundamental
to study the spatial extent of the chemical properties of the ISM
in extended objects. In this context, this technique along with the use of
appropriate statistical tools allows us to explore the presence of chemical inhomogeneities
in both O/H and N/O, and to relate this local
pollution with the position of WR stars, whose stellar winds enrich the
surrounding ISM with the products of the main sequence nuclear burning 
of massive stars.


Our statistical method points to a high degree of homogeneity in  O/H across the 
FoV of the
studied galaxies,  implying scales of several kpc, with the only exception
of WR404, where a possible gradient of metallicity is found. This spatial
variation of the metallicity goes from lower values in the brightest region of the
galaxy at the SE, to slightly higher values in the low surface brightness
tail towards the NW. 

Our spatial analysis of N/O reveals that  only WR057 presents values for which 
the homogeneity of this abundance ratio cannot be ruled out. 
In contrast, for all the other cases the conditions assumed to consider 
a homogeneous distribution of N/O are not fulfilled, as very high values of this ratio in certain 
positions of the FoV were measured. 

Although the WR blue bump is detectable in the SDSS spectra of the six
selected galaxies, we only found it in our PMAS observations in two 
spaxels of the WR039 galaxy. These are
marked with red crosses in Figure \ref{Halfa}. A thorough analysis of the
causes of the missing detection of the bump in the other objects will be 
performed in a forthcoming paper. 
Figure \ref{wrbb} shows a portion of the spectrum obtained by co-adding the emission from the 2 spaxels with
 the WR bump detection. The measured luminosity of
the bump at the adopted distance, once the emission lines over the
bump were removed,  [L$_{\textrm{WR}}$ = 10$^{40.08\pm0.08}$ erg/s],
is consistent with the value measured in the integrated SDSS spectrum 
[L$_{\textrm{WR}}$ = 10$^{39.97\pm0.15}$ erg/s].
In contrast, the equivalent width of the bump in these two spaxels (20$\pm$4 \AA)
is much higher than in the SDSS spectrum (8$\pm$4 \AA), as expected
taking into account the fact that the collected stellar continuum 
in the area of the two 1'' PMAS spaxels is fainter than in the 3'' SDSS fiber.

To investigate the possible connection
between the detected nitrogen overabundance and the location of the WR stars, 
we identified in the histograms shown in Figure \ref{oh} and \ref{no} the
probable positions of the WR stars. In the case of WR039, where the WR bump was detected
in our IFS data, the corresponding
spaxels with WR emission are plotted as red bars. In the rest of the objects,
we selected the four spaxels probably encompassed by the SDSS fiber and which
are thought to host the WR population, and we plotted them in the
histograms as black bars. This subset of spaxels always include the brightest 
H$\alpha$ position in the observed galaxies. 
Finally, to study the possible extent of the N pollution,
we also identified the 12 spaxels around these 4 positions of the SDSS fiber 
as brown bars in the histograms.  As can be seen in the histograms and can
also be confirmed by visual inspection of the N/O maps, the nitrogen overabundance
is tightly related to the position of the WRs in WR038, WR039, WR266, and WR505,
although there is not a perfect match between them. This is well illustrated in the
unique case where we identified the WR emission, WR039, where the N overabundance
is slightly displaced in relation to the position of the WR bumps. 
The other galaxy whose N/O is not homogeneous is WR404, but in
this case, this is probably related to the gradient of 
metallicity across its tail detected in the O/H analysis, as no direct relation
between the N overabundance and the positions of the WRs is detected.

The results obtained in this work, in which we find evidence for a local
nitrogen overabundance (in zones of the order of 100 pc around the position 
of the WR bumps) in four out
of the six observed galaxies by means of IFS have the following implications:

- WR stars are possibly the main cause of the overabundance of nitrogen observed in 4 out of
6 observed galaxies
at spatial scales of the order of several hundreds of parsecs detected around the position of these stars,
differing from the objects studied in P\'erez-Montero et al. (2011) where this overabundance
was detected at scales of several kpc. According to the chemical yields of
massive stars from Moll\'a \& Terlevich (2012) presented in Figure 14 of 
P\'erez-Montero et al. (2011) the stellar masses of the ionizing clusters in the
sample of WR galaxies studied here, (all of them around 10$^7$ M$_{\odot}$) 
can produce a N/O excess at distances compatible with the scales at which the N pollution 
has been detected by means of IFS in this work.

- As the local N pollution has not been observed in all the studied objects
in this work and, in those where it was observed, 
it does not show a perfect match with the
positions of the WR bump. 
Although it is necessary to take the limited spatial resolution of our
observations into account, this mismatch could be possibly due to a timescale 
offset between the
lifetime of the WR stars and the mixing of the ejected material with the ISM.
Possibly, the positions with relatively high N/O trace regions where the WR stars were
present. In contrast, the positions where the bumps are detected are tracing the ongoing
star formation regions. This spatial mismatch between N excess and WR positions
has also been observed in other nearby star-forming objects studied by
means of optical IFS ({\em e.g.} NGC5253, Monreal-Ibero
et al. 2012). 
This timescale offset between WR lifetimes and N mixing 
is supported by the results from other previous works
based on IFS data where WR stars were reported, but not the excess
in N/O ({\em e.g.} in this work, WR057 
and WR404, in HS0837+4717 and Mrk930 in P\'erez-Montero et al. 2011, or in two out of the 
three WR clusters detected in Mrk178 (Kehrig et al. 2013).

- The local nitrogen pollution happens while homogeneous values of the
oxygen abundance are found, which could be indicative that oxygen is not 
noticeably present in the winds ejected by the WR stars at this stage. 
Apparently the properties of
these winds (density, velocity, etc ...) favour the mixing of their components
with the surrounding warm ISM. However, the mixing expected later  of the oxygen
ejected during the last stages of the WR phase and the subsequent O-rich SNe explosions
have a timescale much  longer than that of early WR winds ({\em e.g.} Tenorio-Tagle 1996).

- Additional mechanisms other than the enrichment due to WR winds are thought to be responsible
for the nitrogen overabundance in star-forming galaxies. This is the case
of collisional deexcition of O$^+$ in strong shocks associated with mergers
(Raymond 1979) or other 
hydrodynamical processes, such as the infall of pristine gas
(K\"oppen \& Hensler, 2005), which can
at same time reduce the overall metallicity of galaxies and boost the 
star-formation. These processes could be behind the relation between
metallicity and star-formation rate in galaxies. On the contrary, such
a mechanism will not have any influence on the abundance ratio of metals
as N and O (Edmunds 1990). A very suitable tool to identify these
processes and to distinguish them from local pollution, as in the
case of N ejection by WR stars, is the simultaneous analysis of the
relations between stellar mass and
metallicity (MZR) and with N/O (MNOR) .
This was already used by Amor\'\i n et al. (2010) to understand the 
low metallicity combined with N/O ratio much higher than
the values in the plateau of the diagram O/H vs. N/O measured
in {\em green pea} galaxies. In that work the analysis of these galaxies
shows that these objects have the expected N/O for their masses, while they 
have systematically lower metallicities, even though WR stars have been detected
in deep GTC spectra of some of them (Amor\'\i n et al. 2012).

Hence, for WR galaxies, we try to
understand their average observed N/O excess, as reported by
Brinchmann et al. (2008), doing the same analysis. In Figure \ref{SDSS_WR}
is shown the MZR for the star-forming galaxies
of the SDSS, with their stellar masses compiled from the 
Max Planck Institute for Astrophysics-Johns Hopkins University (MPA-JHU)
catalog\footnote{Available at http://www.mpa-garching.mpg.de/SDSS/}
and oxygen abundances calculated using the N2 parameter for emission lines
with S/N larger than 2. The solid red line, as explained in P\'erez-Montero
et al. (2013) is a quadratic fit to the medians for stellar mass bins
of 0.2~dex. The black points are the matches between the WR galaxy catalog
by Brinchmann et al. (2008) and all the other star-forming SDSS galaxies selected from
the MPA/JHU list.
The number of matches obeying the S/N criterion and having a minimum
redshift ($z > 0.02$) to avoid serious aperture effects in the determination of the stellar
mass, as described in P\'erez-Montero et al. (2013), is 254.
In the
right panel of the same figure is shown the MNOR,
as calculated using the N2S2 parameter with the calibration by P\'erez-Montero
\& Contini (2009). As can be appreciated, and contrary to {\em green pea} galaxies,
the WR galaxies are in average in agreement with the metallicities expected
for their stellar mass but, in contrast, have larger N/O ratios, which could
be just local pollution values in the same region covered by the SDSS fiber
and possibly due to the enrichment by WR stars.

\section{Summary}

In this work we presented 3.5 m CAHA - PMAS IFS observations of six 
metal-poor compact WR galaxies selected from the catalog published 
by Brinchmann et al. (2008) in the optical spectral range 3700 - 6850 \AA. Our aim
is to study the connection between the presence of WR stars
and N/O excess as compared with the values predicted by chemical
evolution models at this metallicity regime.

We derived O/H and N/O abundances ratios using
the direct method ({\em i.e.} with the determination of the electron temperature),
or strong-line methods based on \nii 6584 {\AA} emission line, such as
N2, O3N2 (for O/H), and N2O2 (for N/O) with the calibrations provided by
P\'erez-Montero \& Contini (2009), which are consistent with the direct method.
We studied the homogeneity of the spatial distributions of both O/H and
N/O using the same statistical procedure introduced by P\'erez-Montero et al. (2011)
and improved by Kehrig et al. (2013). 

Our results indicate that in all the studied
objects O/H  can be considered as uniform in scales of the order of several kpc,
with the exception of WR404, for which a gradient of O/H is found
in the same direction of a low surface brightness tail. In contrast, N/O
can only be considered as homogeneous in WR057. In four of the 
six studied galaxies (WR038, WR039, WR266, and WR505) we found 
positions associated with or close to the WR stars with N excess in spatial
scales of the order of several hundreds of pc. We discussed that,
according to the models presented by P\'erez-Montero et al. (2011) based
on massive star yields of Moll\'a \& Terlevich (2012), the N excess length scale is consistent
with the distances at which the stellar clusters in these galaxies can enhance the gas-phase 
abundance of N. On the other hand, our analysis of both
the MZR and the MNOR of the WR galaxies of the SDSS WR catalog of
Brinchmann et al. (2008) excludes hydrodynamical effects, such as
metal-poor gas inflows, as the
more frequent cause of the N excess detected in the SDSS galaxies with
a detection of the WR bump.

\begin{acknowledgements}
Based on observations collected at the Centro Astron\'omico Hispano Alem\'an (CAHA) at Calar Alto, operated jointly by the Max-Planck Institut f\"ur Astronomie and the Instituto de Astrof\'\i sica de Andaluc\'\i a (CSIC).
This work has been partially supported by projects
AYA2007-67965-C03-02 and AYA2010-21887-C04-01
of the Spanish National
Plan for Astronomy and Astrophysics. We also thank an anonymous referee for his/her very
thorough revision of this manuscript that has helped to improve it.

\end{acknowledgements}

\end{document}